\documentclass[10pt,journal,compsoc]{IEEEtran}
\ifCLASSOPTIONcompsoc
  \usepackage[nocompress]{cite}
\else
  \usepackage{cite}
\fi
\usepackage[most]{tcolorbox}
\usepackage{tabularx}
\usepackage{url}
\usepackage{framed}
\usepackage{multirow}
\usepackage{balance}
\usepackage{booktabs}
\usepackage{fontawesome}
\usepackage{threeparttable}
\usepackage[figuresright]{rotating}
\usepackage[colorlinks=true,linkcolor=black,citecolor=black,urlcolor=black]{hyperref}
\usepackage{orcidlink}
\usepackage{caption}
\usepackage{subcaption}
\usepackage{tablefootnote}
\usepackage{float}
\usepackage{enumitem}
\usepackage{colortbl}
\usepackage{latexsym} 
\usepackage{tabularx}
\usepackage{amsfonts} 
\usepackage{array}
\usepackage{longtable}
\usepackage{supertabular}
\usepackage{xcolor}
\usepackage{soul}

\usepackage{graphicx}
\definecolor{lightblue}{RGB}{220,230,241}

\definecolor{lightblue}{RGB}{173, 216, 230}
\definecolor{softpink}{RGB}{255, 182, 193}
\definecolor{mygray}{rgb}{0.9,0.9,0.9}

\hypersetup{
    colorlinks=true,
    linkcolor=blue,
    filecolor=magenta,      
    urlcolor=blue,
    pdftitle={Overleaf Example},
    pdfpagemode=FullScreen,
}

\begin{document}
\begin{sloppypar}

\title{How Do OSS Developers \textcolor{black}{Reuse} Architectural Solutions from Q\&A Sites: An Empirical Study}

\author{Musengamana Jean de Dieu~\orcidlink{0009-0000-5956-997X}, Peng Liang~\orcidlink{0000-0002-2056-5346}, and Mojtaba Shahin~\orcidlink{0000-0002-9081-1354}
\IEEEcompsocitemizethanks{
\IEEEcompsocthanksitem Musengamana Jean de Dieu and Peng Liang are with the School of Computer Science, Wuhan University, 430072 Wuhan, China.\\
E-mail: {mjados@outlook.com, liangp@whu.edu.cn}

\IEEEcompsocthanksitem Mojtaba Shahin is with the School of Computing Technologies, RMIT University, Melbourne, Australia.\\
E-mail: mojtaba.shahin@rmit.edu.au}

\thanks{Manuscript received; revised. (Corresponding author: Peng Liang)}}

\markboth{IEEE Transactions on Software Engineering}
{M. J. de Dieu \MakeLowercase{\textit{et al}.}: IEEE Transactions on Software Engineering}

\IEEEtitleabstractindextext{
\begin{abstract} 
Developers \textcolor{black}{reuse} programming-related knowledge (e.g., code snippets) on Q\&A sites (e.g., Stack Overflow) that functionally matches the programming problems they encounter in their development. Despite extensive research on Q\&A sites, being a high-level and important type of development-related knowledge, architectural solutions (e.g., architecture tactics) and their \textcolor{black}{reuse} are rarely explored. To fill this gap, we conducted a mixed-methods study that includes a mining study and a survey study. For the mining study, we mined 984 commits and issues (i.e., 821 commits and 163 issues) from 893 Open-Source Software (OSS) projects on GitHub that explicitly referenced architectural solutions from Stack Overflow (SO) and Software Engineering Stack Exchange (SWESE). For the survey study, we identified practitioners involved in the \textcolor{black}{reuse} of these architectural solutions and surveyed 227 of them to further understand how practitioners \textcolor{black}{reuse} architectural solutions from Q\&A sites in their OSS development. Our main findings are that: (1) OSS practitioners \textcolor{black}{reuse} architectural solutions from Q\&A sites to solve a large variety (15 categories) of architectural problems, wherein \textit{Component design issue}, \textit{Architectural anti-pattern}, and \textit{Security issue} are dominant; (2) Seven categories of architectural solutions from Q\&A sites have been \textcolor{black}{reused} to solve those problems, among which \textit{Architectural refactoring}, \textit{Use of frameworks}, and \textit{Architectural tactic} are the three most \textcolor{black}{reused} architectural solutions; (3) \textcolor{black}{OSS developers often rely on ad hoc ways (e.g., informal, improvised, or unstructured approaches) to \textcolor{black}{reuse} architectural solutions from SO, drawing on personal experience and intuition rather than standardized or systematic practices}; (4) \textcolor{black}{Reusing} architectural solutions from SO comes with a variety of challenges, e.g., OSS practitioners complain that they need to spend significant time to adapt such architectural solutions to address design concerns raised in their OSS development, and it is challenging to \textcolor{black}{reuse} architectural solutions that are not tailored to the design context of their OSS projects. Our findings pave the way for future research directions, including the design and development of approaches and tools (such as IDE plugin tools) to facilitate the \textcolor{black}{reuse} of architectural solutions from Q\&A sites, and could also be used to offer guidelines to practitioners when they contribute architectural solutions to Q\&A sites. \textcolor{black}{Our dataset is publicly available at \url{https://doi.org/10.5281/zenodo.10936098}.}
\end{abstract}

}

\maketitle 
\IEEEdisplaynontitleabstractindextext
\IEEEpeerreviewmaketitle

\section{Introduction} \label{Introduction}
\textcolor{black}{Building the architecture of a software system involves making a series of design decisions that significantly impact a project's success \cite{jansen2005SoftArch, kruchten2006past}. These decisions arise from architectural problems or concerns, typically emerging in the early and late stages of development. Addressing these problems requires strategic foresight, as they dictate the course of the project \cite{SA2012}. Architectural problems involve making critical decisions, such as defining the system’s structure, determining components and their interactions, and selecting appropriate patterns to meet scalability, performance, and maintainability requirements \cite{SA2012}. In contrast, programming problems focus on lower-level issues like algorithms, data structures, and logic errors, which are more localized and less disruptive to the overall architecture, often resolved with minimal effort \cite{SA2012}.}

\textcolor{black}{To address architectural problems, developers often apply architectural solutions including architectural tactics, such as resource cashing, scheduling tactics for performance, heartbeat tactics for availability, and data encryption tactics for security concerns \cite{SA2012}. Once implemented, these architectural solutions become challenging to modify \cite{SA2012}. This distinction makes software architecture more complex than implementation tasks, such as fixing a bug in a Java method, which tends to be less complex than altering an architectural pattern or technology choice. Understanding and solving architectural problems leads to better software design and can mitigate risks like project failure, cost overruns, and operational inefficiencies \cite{SA2012}. Therefore, given the pivotal role of architectural problems and solutions in shaping software systems, this study focuses on exploring architectural problem-solving activities in Open-Source Software (OSS) development.} 

\begin{figure*}[h!]
\centering \includegraphics[width=1.0\linewidth]{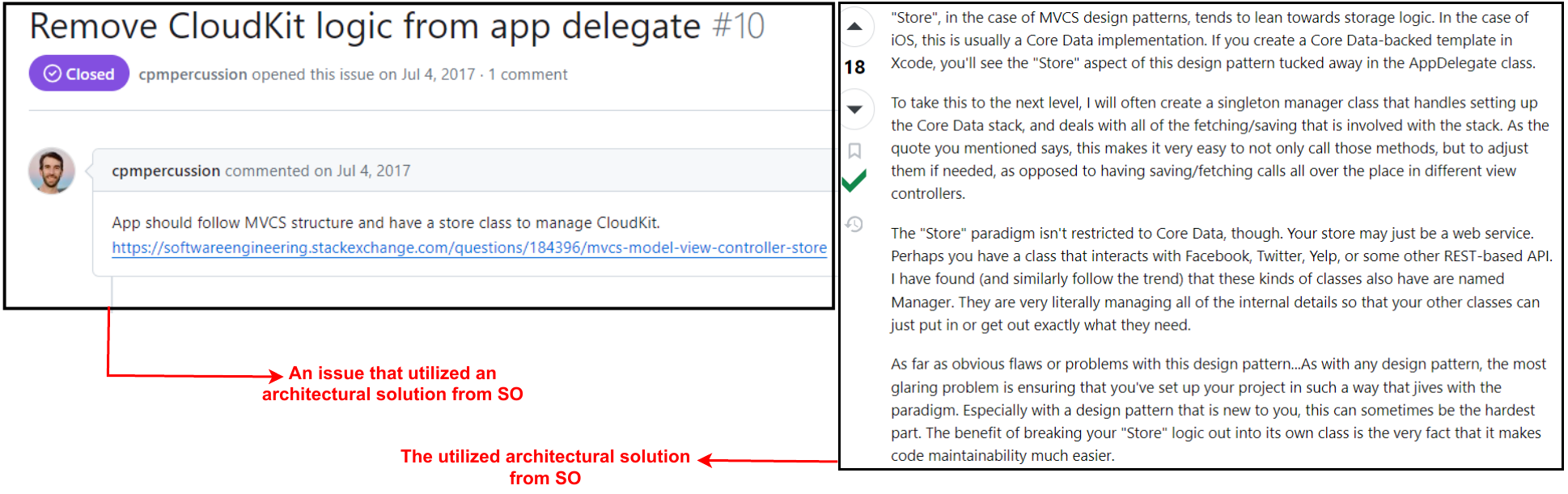} \caption{An issue from GitHub that \textcolor{black}{reuses} an architectural solution from SO to structure the application employing the \textit{Model-View-Controller-Store (MVCS)} architectural pattern in the \textit{microjam} project.} \label{ULRLinktoCommitMessage} 
\end{figure*}

\textcolor{black}{In recent years, real-world software projects have increasingly \textcolor{black}{reused} development knowledge shared on Q\&A sites, as evidenced by many studies (e.g., \cite{lotter2018code, chen2024empirical, huang2022towards, abdalkareem2017code}). However, existing research has primarily focused on the \textcolor{black}{reuse} of development knowledge at a low level of abstraction, such as code-level solutions, rather than at higher levels, like architectural solutions \cite{bi2021mat}. Architectural knowledge shared on Q\&A sites, such as SO, is also \textcolor{black}{reused} to address development issues in real-world projects. For example, Figure \ref{ULRLinktoCommitMessage} illustrates an example of how an architectural solution from SO was \textcolor{black}{reused} in a GitHub project. Specifically, a developer referred to an architectural solution on SO to structure the application using the Model-View-Controller-Store (MVCS) pattern for the \textit{microjam} mobile app. Despite such practical uses, the \textcolor{black}{reuse} of architectural solutions from Q\&A sites in OSS development remains underexplored. This gap limits our understanding of how developers adopt and \textcolor{black}{reuse} these solutions into OSS projects, as well as the challenges they face, such as difficulties in identifying architectural solutions from Q\&A sites for specific design concerns.} 

This \textbf{motivated} us to conduct a study aimed at bridging this gap by investigating the activities involved in \textcolor{black}{reusing} architectural solutions from Q\&A sites, particularly SO, in OSS development. Specifically, this study explores how these solutions are adopted and \textcolor{black}{reused} into real-world OSS projects, uncovering insights into various aspects, including the challenges practitioners face. Our aim is to provide guidance to researchers and practitioners, contributing to the development of approaches, tools, documentation, and guidelines that improve the adoption and integration of solutions from SO into OSS development. \textcolor{black}{Below, we present the goal of this study.} 

\textcolor{black}{The \textbf{goal} of this study is to \textit{analyze} the current practices of reusing architectural solutions from SO in OSS development, for the \textit{purpose of} understanding and gaining deeper insights into the activities involved. Specifically, the study aims to: (1) investigate the architectural problems solved and the architectural solutions reused, and (2) explore the ways, characteristics, and challenges practitioners face when reusing architectural solutions from SO, \textit{from the point of view of} practitioners \textit{in the context of} OSS development.} 

This study is \textbf{important} for several reasons: (i) Architectural problems in OSS development can greatly affect software quality, including performance, scalability, and maintainability. Since different types of architectural issues often require distinct strategies, researching and categorizing these issues is essential to identify the specific design problems faced by OSS developers, enabling the creation of more relevant solutions and targeted solutions. (ii) This study highlights recurring architectural problems and common solutions from SO in OSS projects, which can inform the creation of better documentation, approaches, and tools that integrate architectural solutions from SO more effectively into OSS workflows. (iii) Many OSS contributors rely on community resources like SO for architectural guidance \cite{soliman2016architectural}. Researching this reliance provides insights into the challenges architects and developers face when \textcolor{black}{reusing} architectural solutions shared on these sites. This understanding can help users of Q\&A sites present architectural knowledge in better-structured ways. Additionally, such insights can guide Q\&A site maintainers in creating structured guidelines for contributors, promoting clearer and more reusable architectural solutions. Ultimately, this fosters a more effective architectural knowledge-sharing environment, enabling developers to more easily find and apply architectural solutions in their projects.

To \textbf{conduct this study}, we were guided by guidelines for selecting the empirical methods for software engineering research \cite{easterbrook2008selecting} and chose a mixed-methods approach (i.e., a mining study and a survey study). This involved collecting data from two sources: GitHub and a survey. Specifically, we began with a mining study to identify commits and issues that referenced and \textcolor{black}{reused} architectural solutions from Q\&A sites. We extracted a total of 984 commits and issues (comprising 821 commits and 163 issues) from 893 publicly available OSS projects on GitHub, each with at least five contributors, covering diverse domains such as development frameworks, games, and networking and communications. These commits and issues referenced architectural solutions from SO or Software Engineering Stack Exchange (SWESE), hereafter referred to as ``SO posts'' for simplicity. The data collection period for these commits and issues spanned from September 15, 2008 (the official launch date of SO) to February 25, 2024 (the start of this study). Additionally, we identified practitioners involved in \textcolor{black}{reusing} these architectural solutions from the commits and issues. We reached out to them via email, providing a link to the project and a link to the survey questionnaire, inviting them to participate. The survey was conducted online and comprised both closed-ended and semi-closed questions. The closed-ended questions included a fixed set of multiple-choice items and numerical values. The semi-closed questions offered predefined options or categories (e.g., categories of architectural problems, derived from the mining study) for participants to select, while also allowing them to provide additional responses in an “Other” field. In total, we received responses from 227 practitioners, which we analyzed to gain a deeper understanding of how they \textcolor{black}{reuse} architectural solutions from SO posts in their OSS development. 

\textcolor{black}{To achieve the goals of this study, we formulated five \textbf{Research Questions} (RQs). The first three RQs (RQ1, RQ2, and RQ3) are answered using data collected from both the mining study and the survey (responses from practitioners). The remaining two RQs (RQ4 and RQ5) are particularly addressed through the survey. The research questions and their rationale are explained as follows:}

\color{black}
\textbf{RQ1. What architectural problems in OSS development are solved by \textcolor{black}{reusing} architectural solutions from SO?}

\textbf{Rationale}: Developers encounter various architectural problems in OSS development, such as component design issues, architectural anti-patterns, and performance bottlenecks. To address these problems, they frequently turn to SO for architectural solutions. This RQ aims to identify the types of architectural problems OSS developers encounter and resolve \textcolor{black}{reusing} SO's architectural solutions. Answering this RQ can inspire researchers to delve deeper into these problems and equip developers with knowledge about the architectural issues that can be effectively solved through SO' solutions.

\textbf{RQ2. What architectural solutions from SO are \textcolor{black}{reused} by OSS developers?}

\textbf{Rationale}: Developers often \textcolor{black}{reuse} various types of architectural solutions from SO, including architectural tactics (e.g., heartbeat \cite{SA2012}) and architectural patterns (e.g., authenticator pattern \cite{steinegger2016risk}), for purposes such as enhancing system reliability and security. This RQ aims to identify the types of architectural solutions from SO that OSS developers adopt and \textcolor{black}{reuse} in their projects. The answer to this RQ can help other interested developers know and \textcolor{black}{reuse} these architectural solutions in the same development contexts. Furthermore, researchers may want to investigate the efficacy and adaptability of these solutions in diverse development scenarios, ultimately contributing to a broader understanding of architectural practices in software engineering.

\textbf{RQ3. \textcolor{black}{How do OSS developers reuse architectural solutions from SO in their projects?}} 

\textbf{Rationale}: \textcolor{black}{Previous research has focused on proposing methods for \textcolor{black}{reusing} development knowledge, particularly source code from SO (\cite{ponzanelli2014mining, ponzanelli2013leveraging}). For example, Ponzanelli \textit{et al}. \cite{ponzanelli2013leveraging} developed a tool to generate context-based queries in the Eclipse IDE, retrieving relevant code and discussions from SO. However, no study has systematically examined how developers \textcolor{black}{reuse} development knowledge, including architectural solutions, from their own perspectives. Without such insights, automated approaches and tools to support the \textcolor{black}{reuse} of architectural solutions risk being driven by researchers' assumptions, which may not align with developers' needs. This RQ aims to uncover how (e.g., in what ways) developers \textcolor{black}{reuse} architectural solutions from SO in OSS development, offering actionable guidance for practitioners and motivating the creation of more aligned approaches and tools that support the effective integration and reuse of SO-based architectural solutions.}

\textbf{\textcolor{black}{RQ4. What characteristics of architectural solutions and their descriptions on SO do OSS developers consider important for reusing those solutions in development?}}

\textbf{Rationale}: \textcolor{black}{The exploration of architectural solutions on SO has yielded a wealth of knowledge and practices that OSS developers can leverage \cite{soliman2016architectural, soliman2017developing, soliman2021exploring}. However, the effectiveness and applicability of these solutions can significantly depend on their specific features. \textcolor{black}{This RQ aims to provide insights on what specific features of the architectural solutions and their descriptions on SO make them more reusable or helpful in OSS development.} 
 Such insights can help researchers shed light on the criteria employed by OSS developers when selecting and \textcolor{black}{reusing} architectural solutions from SO, so that they can further explore these features. Additionally, these insights can guide other developers interested in evaluating architectural solutions from SO that best align with their project requirements, ultimately enhancing the quality of their systems. Moreover, SO owners and users can use these insights to enhance how architectural solutions are shared on the platform.} 
 
\textbf{RQ5. What challenges do OSS developers face when \textcolor{black}{reusing} architectural solutions from SO?}

\textbf{Rationale}: \textcolor{black}{Developers may face challenges when \textcolor{black}{reusing} architectural solutions from SO in OSS development, such as assessing their relevance or up-to-dateness (e.g., alignment with the latest technologies). This RQ seeks to uncover these challenges, offering researchers insights into the complexities developers face. Understanding these difficulties can help researchers devise strategies and approaches to better support OSS developers in effectively \textcolor{black}{reusing} SO-based architectural solutions.} 

In summary, the \textbf{contributions} made by this paper are:

\begin{enumerate}[leftmargin=3ex]
\item A comprehensive list of architectural problems encountered in OSS development, along with the corresponding architectural solutions from SO \textcolor{black}{reused} to address them. We provided the mapping relationship between the encountered architectural problems and solutions reused. This provides valuable insights for architects and developers, enabling them to identify and reuse these solutions in similar development contexts effectively.
\item A curated list of five distinct methods OSS developers employ to \textcolor{black}{reuse} architectural solutions from SO, highlighting clear patterns in their engagement with these solutions. This list serves as a valuable resource for developers looking to adopt similar approaches effectively. 
\item \textcolor{black}{A comprehensive list of nine characteristics OSS developers prioritize when reusing architectural solutions from SO. These characteristics provide researchers with insights into the criteria used by OSS developers when adopting architectural solutions from SO, paving the way for further exploration and deeper understanding of these selection factors.}
\item \textcolor{black}{A list of challenges developers face when reusing architectural solutions from SO in OSS development. These challenges provide insights to inspire the development of approaches and tools for adapting and reusing architectural solutions from Q\&A sites, such as SO.}
\end{enumerate}
\color{black}

\textcolor{black}{The remainder of this paper is structured as follows. Section \ref{RelatedWork} describes the related work. Section \ref{Methodology} details the research methodology employed. Section \ref{Results} provides the study results, which are further discussed in Section \ref{Implications}. The potential threats to validity are clarified in Section \ref{ThreatsValidity} and Section \ref{Conclusions} concludes this work with future directions.}

\section{Related Work}\label{RelatedWork}
\textcolor{black}{This section reviews related work relevant to our study, focusing on two main themes: \textit{architectural knowledge in Q\&A sites and software respositories} (Section~\ref{ArchitecturalKnowledge_In_QA_Sites}) and \textit{code reuse from Q\&A sites} (Section~\ref{Code_Reuse_from_QA}). We then provide a structured comparison of each theme with our work (Section~\ref{Conclusive_Summary}).}

\subsection{Architectural Knowledge in Q\&A Sites \textcolor{black}{and Software Repositories}}\label{ArchitecturalKnowledge_In_QA_Sites} \textcolor{black}{\textbf{Architectural issues}, such as inter-component communication issues, cyclic dependencies \cite{feng2024empirical}, modularity violations \cite{baldwin2000design}, and architectural deployment issues, arise during software architecture design due to the complexity of decisions regarding component structuring, interface definition, and interaction management \cite{SA2012}. These issues often involve balancing competing requirements like performance, scalability, and security while adhering to constraints such as technology limitations, budgets, and timelines \cite{SA2012}. They may also stem from challenges in managing dependencies, ensuring modularity, and resolving trade-offs between flexibility and maintainability. Architectural issues can emerge during the design phase or post-deployment, requiring careful resolution to sustain or improve software quality}. 

Several studies have investigated architectural issue provided in Architectural Related Posts (ARPs) on SO from different perspectives. For instance, Raida \textit{et al}. leveraged SO to classify questions, including architectural questions or issues, based on their difficulty using contextual features. Their findings indicated that architectural questions are notably among the more challenging problems on the platform \cite{raida2024study}. Soliman \textit{et al}. \cite{soliman2016architectural} analyzed architectural-related posts (ARPs), including architectural questions, that discuss technology decisions on SO  \cite{kruchten2004ontology}. They classified ARPs across two dimensions: the \textit{purpose of the question} (solution synthesis, solution evaluation, and multi-purpose inquiries) and the \textit{solution type} (technology feature, technology bundle, and architecture configuration), identifying six distinct ARP categories. In a subsequent study, Soliman \textit{et al}. \cite{soliman2017developing} developed an ontology to cover Architectural Knowledge (AK) concepts found in SO. Soliman \textit{et al}. \cite{soliman2018improving} introduced a domain-specific search method for retrieving AK, including architectural issues, from SO, demonstrating that this new method outperforms conventional keyword-based search approaches like Google. Sul{'\i}r and Regeci identified common development questions on SO, including architectural problems, emphasizing topics like database systems, quality assurance, and agile development. Tian \textit{et al}. explored architectural smells (ASs) on SO, revealing a lack of specialized tools for their refactoring and noting that ASs arise from pattern violations, design principle misuse, and antipattern application \cite{tian2019developers}. Kozanidis \textit{et al}. \cite{kozanidis2022asking} analyzed technical debt (TD)-related questions on SO, including architecture debt, highlighting the challenges developers face in addressing TD and emphasizing the importance of early intervention to mitigate its long-term effects.  

\textbf{Architectural solutions}, including patterns and tactics, are part of the high-level artifacts and core elements of software design. They address concerns like performance and reliability of the system \cite{SA2012}. A few existing studies have examined architectural solutions shared on SO including architectural tactics and patterns. Bi \textit{et al}. \cite{bi2021mat} used a semi-automatic mining approach to extract discussions on Architectural Tactics (ATs) and Quality Attributes (QAs), creating a knowledge base to guide AT-related design decisions. Chinnappan \textit{et al}. \cite{chinnappan2021architectural} mined energy-efficient robotics tactics from SO and other repositories, presenting them in a generic, UML-inspired format for broader applicability. Similarly, Malavolta \textit{et al}. \cite{malavolta2021mining} analyzed data from multiple sources, including SO, to identify green architectural tactics within the ROS ecosystem. They conducted an empirical evaluation to assess these tactics' effectiveness in enhancing energy efficiency and sustainability. Wijerathna \textit{et al}. \cite{wijerathna2022mining} examined the relationship between design contexts and patterns on SO, introducing a mapping framework to match patterns with contexts based on developer discussions. \textcolor{black}{Aktar \textit{et al}.~\cite{aktar2025architecture} extracted SO posts and GitHub issues to investigate architectural decisions as well as the challenges developers face when making such decisions in quantum software development. Soliman \textit{et al}. \cite{soliman2025large} conducted an exploratory case study with 14 software engineers who queried GPT about AK in an open-source system to evaluate LLMs' ability to retrieve embedded AK. They found that GPT demonstrated higher recall than precision. In another study, Soliman \textit{et al}. \cite{soliman2024exploring} analyzed architectural decisions in mailing lists and issue trackers, examining their types, co-occurrences, and discussion methods. They also applied similarity algorithms to link related ADDs across emails and issues, supporting empirically grounded approaches for identifying and classifying ADDs in software design.}

\subsection{Code Reuse from Q\&A Sites}\label{Code_Reuse_from_QA}
\textcolor{black}{Lotter \textit{et al}. \cite{lotter2018code} analyzed ~150K SO posts, identifying that 3.3\% of cloned content in 12 GitHub Java projects originated from SO, while 77.2\% came from reuse among GitHub projects. Abdalkareem \textit{et al}. \cite{abdalkareem2017code} analyzed code reuse from SO in mobile applications and found that 1.3\% of mobile apps in their dataset reused SO code, often integrated late in development. Wu \textit{et al}. \cite{wu2019developers} studied how developers use SO-sourced code in OSS. They found that 31.5\% of files required modifying the code, including solutions from non-accepted answers. Yang \textit{et al}. \cite{yang2017stack} analyzed SO's influence on GitHub by studying shared code snippets. They found 86\% of Python GitHub snippets originated from other GitHub content, with little duplication from SO. Ponzanelli \textit{et al}. introduced tools like SEAHAWK and Prompter to assist developers in finding and integrating SO code into IDEs \cite{ponzanelli2014mining}. Chen \textit{et al}. \cite{chen2024empirical} studied programmers' code reuse, comparing open-source project commits with SO snippets. They found a 6.32\% average SO code reuse across 793 Java projects, with experienced developers more inclined to reuse SO snippets.}

\begin{table*}[!h]
\footnotesize
\caption{\textcolor{black}{Comparison between our findings and prior findings}}
\label{ComparisonBetweenOurFindingsandPriorFindings}
\begin{tabular}{|m{1.6cm}|m{5.1cm}|m{3.9cm}|m{5.9cm}|}
\hline
\textcolor{black}{\textbf{Theme}}                                          & \textcolor{black}{\textbf{Our Study}}           & \textcolor{black}{\textbf{Prior Study}}  & \textcolor{black}{\textbf{Comparison}} \\\hline

\textcolor{black}{\textbf{Architectural Knowledge in Q\&A Sites \textcolor{black}{and Software Repositories}}}
                                                        & \textcolor{black}{\textbf{Analyzing categories of architectural problems in Q\&A sites}. OSS developers face a range of architectural issues, including \textit{Component design issues}, \textit{Component communication issues}, \textit{Architectural anti-pattern}, and \textit{Performance issues}, and often turn to SO for architectural solutions to address these problems.} 
                                                        
                                                        & \textcolor{black}{Tian \textit{et al}. \cite{tian2019developers} found that SO users discuss architectural smells. Raida \textit{et al}. \cite{raida2024study} found that architectural questions rank among the most challenging on SO. Soliman et al. \cite{soliman2016architectural, soliman2018improving} classified ARPs on SO into categories like technology  evaluation.} 
                                                        
                                                        & \textcolor{black}{Unlike previous studies \cite{tian2019developers, raida2024study, soliman2016architectural, soliman2018improving}, which focused on architectural problems discussed on Q\&A sites, our study investigates these problems within the context of real-world OSS projects. Additionally, we examines how these problems are solved in practice through the \textcolor{black}{reuse} of architectural solutions from SO.}\\\cline{2-4}

                                                        & \textcolor{black}{\textbf{Analyzing categories of architectural solutions in Q\&A sites}. Developers \textcolor{black}{reuse} distinct architectural solutions from SO, such as \textit{Architectural refactoring}, \textit{use of frameworks}, \textit{architectural tactics}, and \textit{Use of APIs}, to solve different architectural problems in OSS development.} 
                                                        
                                                        &\textcolor{black}{Prior studies proposed approaches to extract architectural solutions like architectural tactics \cite{bi2021mat, chinnappan2021architectural, malavolta2021mining, soliman2021exploring}, architectural patterns \cite{wijerathna2022mining, soliman2021exploring, zalewski2021supporting}, frameworks \cite{soliman2016architectural, zalewski2021supporting, karthik2019automatic}, and APIs \cite{zalewski2021supporting} from SO.}
                                                        
                                                        & \textcolor{black}{Our study confirms and aligns with prior findings (e.g., \cite{bi2021mat, chinnappan2021architectural, malavolta2021mining, soliman2021exploring, wijerathna2022mining, soliman2016architectural, zalewski2021supporting, aktar2025architecture}) that SO contains architectural solutions. While previous research primarily focuses on methods for extracting these solutions, our work explores their practical application in real-world OSS projects.}  \\\cline{2-4}
                                                        
                                                        &\textcolor{black}{\textbf{Analyzing characteristics of architectural solutions in Q\&A sites}. Practitioners consider several key characteristics of architectural solutions from SO when \textcolor{black}{reusing} these solutions in OSS development, such as \textit{the architectural solution description includes a sample of code} and \textit{the architectural solution description provides the design context}.
                                                        } 
                                                        
                                                        & \textcolor{black}{Previous studies characterized SO posts from different perspective, such as community factors (e.g., is-accepted, answer score), answer/question factors (e.g., line of code, ages, text lengths, code complexity, comprehensiveness), and user factors (e.g., interest) \cite{geremia2019characterizing, kozanidis2022asking, kabir2024stack, asaduzzaman2013answering, chen2024developers, saha2013toward, aly2021practitioners}.} 
                                                        
                                                        & \textcolor{black}{Prior studies \cite{geremia2019characterizing, kozanidis2022asking, kabir2024stack, asaduzzaman2013answering, chen2024developers, saha2013toward, aly2021practitioners}, predominantly characterized SO posts without situating their analyses within specific development contexts. In contrast, our study adopts a context-aware perspective, focusing on the characteristics of architectural solutions and their descriptions on SO that OSS developers consider when reusing those solutions. By doing so, we identify the specific features of SO posts (i.e., architectural solutions) that developers prioritize when selecting and \textcolor{black}{reusing} SO knowledge into real-world OSS projects. These findings contribute valuable insights to both academic research and the design of practical tools tailored to developers’ workflows.} \\\cline{1-4} 
\textcolor{black}{\textbf{Development Knowledge Reuse from Q\&A Sites}}
                                                        & \textcolor{black}{OSS developers often rely on ad hoc ways (e.g., informal, improvised, or unstructured approaches) to \textcolor{black}{reuse} architectural solutions from SO, drawing on personal experience and intuition rather than standardized or systematic practices. This underscores a lack of formalized methodologies or tools for integrating these solutions into OSS projects. Challenges include the time needed to adapt solutions and mismatched design contexts, highlighting areas for improvement in the \textcolor{black}{reuse} of SO-sourced architectural solutions.}
                                                        
                                                        & \textcolor{black}{Both manual and automatic approaches have been employed to investigate the the code reuse from SO in OSS development from different perspective  \cite{lotter2018code,yang2017stack,guizani2021long, dawood2019mapping,butler2022considerations, ajigini2014towards, balali2018newcomers, dann2021identifying, canfora2020investigating, ayala2018system, gousios2016work, bacchelli2013expectations,joorabchi2013real, chen2008empirical, huang2022towards, ajayi2019toward, wessel2018power, zou2019smart, wu2019developers, chen2024empirical, abdalkareem2017code}.}
                                                        
                                                        & \textcolor{black}{Unlike previous studies \cite{lotter2018code,yang2017stack,guizani2021long, dawood2019mapping,butler2022considerations, ajigini2014towards, balali2018newcomers, dann2021identifying, canfora2020investigating, ayala2018system, gousios2016work, bacchelli2013expectations,joorabchi2013real, chen2008empirical, huang2022towards, ajayi2019toward, wessel2018power, zou2019smart, wu2019developers, chen2024empirical, abdalkareem2017code}, which predominantly analyzed low-level development knowledge such as source code, our study broadens this research landscape by focusing on the distinct practices and challenges OSS developers encounter when \textcolor{black}{reusing} and integrating architectural solutions from SO into their projects.} \\\cline{1-4} 
\end{tabular}  
\end{table*}

\subsection{Conclusive Summary}\label{Conclusive_Summary}
Table \ref{ComparisonBetweenOurFindingsandPriorFindings} highlights the key differences between our study findings and prior research findings. Previous studies (e.g.,~\cite{raida2024study, tian2019developers, soliman2017developing, raida2024study}) primarily focused on extracting textual AK from Q\&A sites like SO to understand its role in software development. In contrast, our study delves into the practical \textcolor{black}{reuse} of architectural solutions sourced from SO in real-world systems, particularly in OSS. Grounded in developers' perspectives, we examine critical aspects such as the architectural problems they address, the solutions they \textcolor{black}{reuse}, and the challenges encountered. Moreover, unlike previous works (e.g.,~\cite{guizani2021long, dawood2019mapping, butler2022considerations, ajigini2014towards, balali2018newcomers}) that focus on low-level code reuse, our research centers on high-level artifacts, emphasizing architectural solutions and their practical implications. By analyzing data from GitHub and a survey, we provide comprehensive insights for researchers, architects, developers, tool designers, and SO stakeholders. Our study not only complements prior work by bridging the gap between extracting AK and understanding its application but also fills a crucial void in knowledge regarding the adoption and utility of architectural solutions from Q\&A like SO in practice.

\section{Methodology} \label{Methodology}
\textcolor{black}{This section presents our mixed-methods study, which consists of a mining study (Section~\ref{Mining_Study}) and a survey study (Section~\ref{Survey_Study}). We begin by explaining the overall research design (Section~\ref{Overall_Research_Design}), including how the two methods complement each other in addressing the RQs introduced in Section~\ref{Introduction}.}

\subsection{Overall Research Design}\label{Overall_Research_Design}

\begin{figure*}[htbp]
 \centering
 \includegraphics[width=1\linewidth]{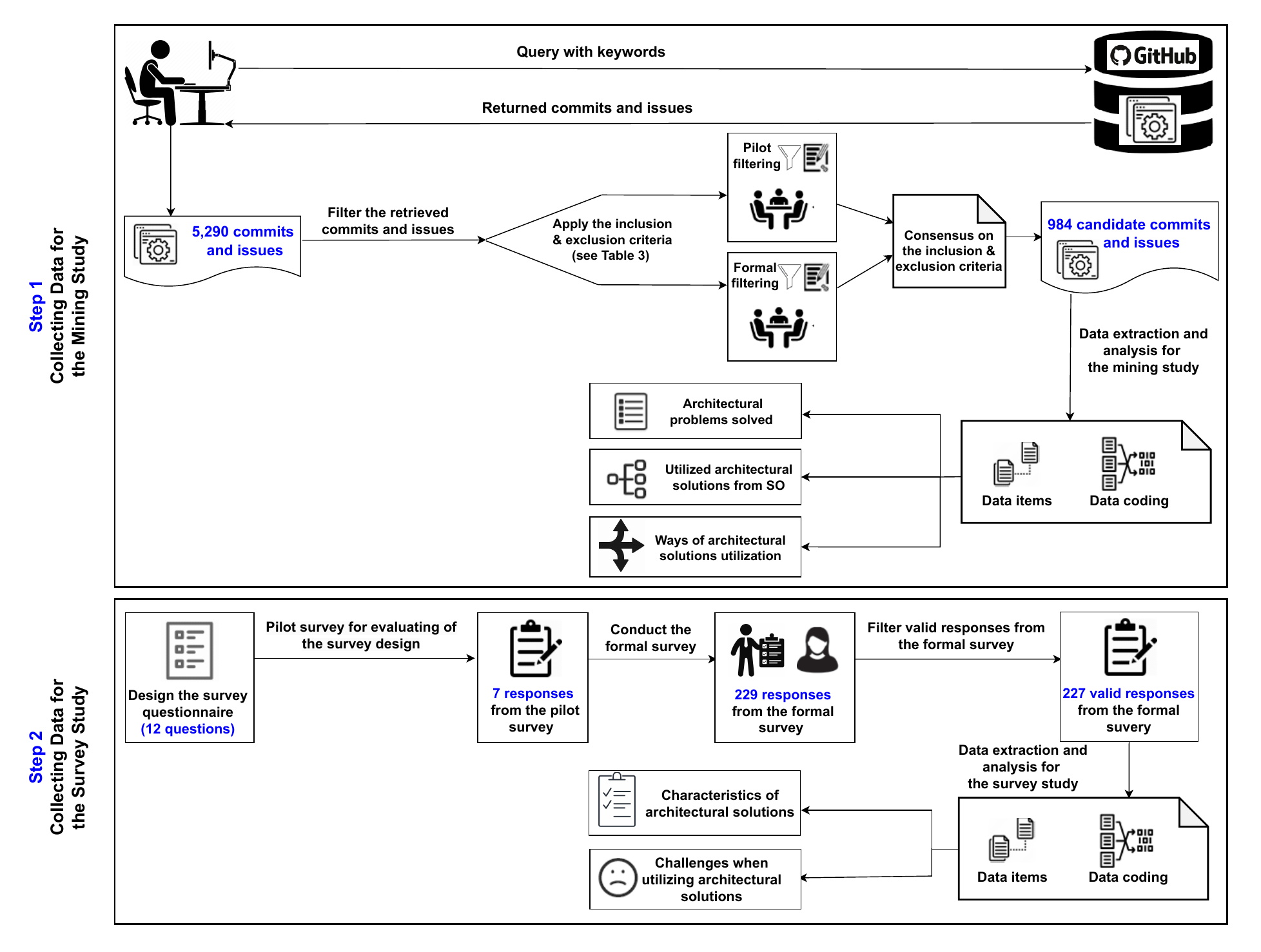}
 	\caption{Overview of the research process}
 \label{OverviewOftheStudyProcess}
\end{figure*}

To conduct this study, we developed a comprehensive research protocol outlining the objectives, methodology, data collection, filtering procedures, and analysis plan (e.g., qualitative analysis and cross-validation among researchers). As detailed in Section \ref{Introduction}, we formulated five research questions (RQs). RQ1–RQ3 are addressed using data from the mining study (GitHub commits and issues) and the survey (practitioner responses), while RQ4 and RQ5 focus solely on survey data. Adopting a mixed-methods approach, we combined GitHub and survey data, as guided by software engineering research methods \cite{easterbrook2008selecting}. This approach was motivated by our pilot study, which revealed that OSS developers rarely addressed aspects relevant to RQ4 and RQ5 directly in commit messages or issue comments. Similar to the study by Li \textit{et al}. \cite{li2020qualitative}, our survey was designed not to validate but to complement the results of the mining study. \textcolor{black}{Although commits and issues lacked explicit statements related to RQ4 and RQ5, they offered indirect clues about potentially relevant factors. To develop response options for RQ4 and RQ5, we drew upon three sources: (1) observations from our pilot study of architectural solution reuse from SO in commit files and issue threads; (2) feedback from two experienced researchers in software architecture and two seasoned OSS developers, who helped validate and refine our draft survey; and (3) insights from our prior work on mining architectural information from SO~\cite{de2023characterizing} and understanding developers' architectural information-seeking behaviors~\cite{de2022developers}}. These survey insights enriched the mining study results, as discussed further in Section~\ref{Survey_Study}. Figure \ref{OverviewOftheStudyProcess} provides an overview of the research process used to answer the RQs elaborated in the Introduction section (see Section \ref{Introduction}).

\color{black}
\subsection{Mining Study}\label{Mining_Study}
\subsubsection{Collecting Data for the Mining Study}\label{CollectingDataforMiningStudy}
\color{black}
\textbf{Searching for potential candidate commits and issues}. We collected commit messages and issue comments from GitHub that explicitly reference SO. \textcolor{black}{Commits capture changes to the codebase, including a unique identifier, message, and metadata like the author and date, providing a historical record and enabling collaboration. Issues, on the other hand, track tasks, bugs, and feature requests and are used to organize and prioritize work according to their priorities. They are often linked to commits to show how changes address specific tasks \cite{chacon2014pro}.}

Developers must adhere to certain licenses (e.g., MIT, CC BY-SA 3.0) when reusing answers or solutions from Q\&A sites. For instance, developers may copy source code from SO posts into their own projects as long as they comply with the CC BY-SA 3.0 license requirements\footnote{\url{https://creativecommons.org/licenses/by-sa/3.0/}}, as specified in an official SO blog post\footnote{\url{https://stackoverflow.blog/2009/06/25/attribution-required/}}. A key requirement of this license is attribution, which developers fulfill by linking to the original SO post in their project files. Several studies (e.g., \cite{huang2022towards, manes2021studying, ragkhitwetsagul2019toxic, chen2024empirical}) have explored code reuse from SO in OSS projects by analyzing GitHub commits. These commits may document bug fixes (e.g., coding or architectural issues) or new feature additions, often reflecting design or coding challenges encountered during development. Building on these studies, we investigated the \textcolor{black}{reuse} of architectural solutions from SO in OSS development through commit analysis. In addition, we analyzed closed issues, specifically excluding open issues, as open issues lack accepted, confirmed solutions. To identify commits and closed issues referencing architectural solutions from SO posts (both questions and answers), we searched for entries containing at least one link to an SO post. \textcolor{black}{We extracted commits and issues from publicly available OSS projects, each with at least five contributors, spanning diverse domains such as development frameworks, games, and networking and communications, using GitHub's advanced search feature\footnote{\url{https://github.com/search/advanced}}. The data collection period for these commits and issues spanned from September 15, 2008 (the official launch date of SO) to February 25, 2024 (the start of this study).}

Before the formal search and filtering processes, we first conducted a pilot search and filtering process to formulate the search terms. We performed a pilot search with several terms, namely ``stackoverflow'' AND ``architect*'', ``stackoverflow'' AND ``design*'', ``softwareengineering.stackexchange'' AND ``architect*'', and ``softwareengineering.stackexchange'' AND ``design*'' to search into commit messages and issue comments through the GitHub search main page. After the pilot search with the mentioned terms, we found that GitHub practitioners mostly use the terms ``design*'' (i.e., ``design'' and ``designing'') in commit messages or issue comments to refer to programming context, for instance, a Uint32Array design\footnote{\url{https://github.com/saghul/txiki.js/issues/369}}. Therefore, we selected the terms ``stackoverflow'' AND ``architect*'' and ``softwareengineering.stackexchange'' AND ``architect*'' to be used in our search. The search terms that were utilized and the number of commits and issues that were obtained from GitHub are reported in Table \ref{SearchTerms}. The search process resulted in 5,290 commits and issues (see Figure~\ref{OverviewOftheStudyProcess}). 

\begin{table*} [h!]
\footnotesize
\caption{Search terms used for mining potential candidate commits and issues from GitHub}
\label{SearchTerms}
\begin{tabular}{m{0.22cm}<{\centering}m{2.3cm}m{8.5cm}<{\centering}m{2.5cm}m{2.5cm}}
\hline
\textbf{\#}  & \textbf{General search terms} & \textbf{Full set of search terms} & \textbf{Num. of retrieved commits \& issues}  & \textbf{Num. of selected commits \& issues}          \\ \midrule
ST1         &  ``StackOverflow'' and ``architect*''                 & ``stackoverflow AND architect'',  
                                                                ``stackoverflow AND architecture'', 
                                                                ``stackoverflow AND architectural'', 
                                                                 and ``stackoverflow AND architecting''                     & 4,316 & 920 (i.e., 810 commits and 110 issues) \\\hline
ST2         &  ``softwareengineering.\newline stackexchange''
               and ``architect*''                                 & ``softwareengineering.stackexchange AND architect'', 
                                                                ``softwareengineering.stackexchange AND architecture'', 
                                                                ``softwareengineering.stackexchange AND architectural'', 
                                                                 and ``softwareengineering.stackexchange AND architecting''   & 974 & 64 (i.e., 11 commits and 53 issues)\\\hline
\textbf{Total}         &                                         &                                                          & 5,290 & 984 (i.e., 821 commits and 163 issues) \\\hline                                                                  
\end{tabular}
\end{table*}

\textbf{Filtering the retrieved commits and issues}. \textcolor{black}{The filtering process involved collaboration: while the first author conducted the pilot and formal data filtering, the results were reviewed and validated by the other two authors. Specifically, before the formal filtering of commits and issues, a pilot filtering was performed whereby the first author randomly selected 10 commits and 10 issues from the 5,290 commits and issues. He manually checked them and drafted a list of inclusion and exclusion criteria.} The other two authors checked and examined the results so that all the authors (three authors) of this study could reach a consensus on the understanding of the defined criteria (see Table~\ref{InclusionExclusionCriteria}). \textcolor{black}{The first author continued with the formal filtering of commits and issues based on the inclusion and exclusion criteria defined during the pilot process}. \textcolor{black}{The first author iteratively refined the criteria throughout the former data filtering process, revisiting and adjusting the inclusion/exclusion criteria as the number of commits and issues increased. When commits or issues were unclear, and the first author encountered ambiguity during filtering, meetings with the other two authors were held to discuss and resolve any confusion. This collaborative process helped enhance the consistency and reliability of our filtering criteria}. The process continued till all the 5,290 commits and issues were manually checked. This step resulted in 984 selected commits and issues (i.e., 821 commits and 163 issues), which belong to 893 open-source projects. The results of this round were checked and verified by the other two authors of this study. In the last column of Table \ref{SearchTerms}, we showed the number of selected commits and issues from the search terms.

\begin{table}
\small
\caption{Inclusion and exclusion criteria for filtering the retrieved commits and issues}
\label{InclusionExclusionCriteria}
\begin{tabular}{p{8.3cm}}
\toprule
\textbf{Inclusion criteria}\\
\midrule
\textbf{I1.} A commit or issue should contain a discussion on software architecture, for example, architecture design and architectural pattern.\\
\textbf{I2.} A commit or issue should contain at least one URL link to an SO post (either a question or an answer/architectural solution) and this commit or issue \textcolor{black}{reused} the architectural solution from the SO post.\\
\textbf{I3.} A commit or issue should contain at least one data item that can be extracted according to the data items in Table \ref{RelationshipDataAnaysisMethodandRQandSQ}.\\
\midrule
\textbf {Exclusion criteria}
\\ \midrule
\textbf{E1.} Duplicate commit or issue.\\
\textbf{E2.} A commit or issue of a toy project (e.g., student projects or course assignments) \cite{kalliamvakou2014promises}.\\
\textbf{E3.} A commit or issue without a detailed description.\\
\bottomrule 
\end{tabular}
\end{table}

\color{black}
\subsubsection{Data Extraction and Analysis for the Mining Study}\label{Data_extraction_analysis_for_mining_study}
\color{black}
\textcolor{black}{This process involved at least two authors: the first author initiated the process, while the second author provided oversight and monitoring. To ensure an effective and systematic data extraction, we followed these steps and guidelines: (i) The first author conducted a pilot data extraction, extracting an initial list of data items from a sample of 5 commits and 5 issues. (ii) To maintain consistency and streamline the data review process, the first author recorded the metadata of these commits and issues in the MAXQDA\footnote{\url{https://www.maxqda.com/}} tool, including the URL link, a description, and the data item to be extracted. (iii) For transparency in decision-making, the first author documented the rationale behind each data extraction decision, noting any ambiguities or uncertainties encountered. (iv) The first author held a meeting with the second author to discuss and refine the extraction guidelines (if necessary), address issues, and resolve ambiguities. The first author proceeded with formal data extraction by extracting data from the remaining selected commits and issues. During this process, any ambiguous words, phrases, sentences, or paragraphs were documented for further discussion with the second author. Finally, the second and third authors reexamined the extraction results of all the commits and issues to ensure all data items were extracted correctly}. Table \ref{RelationshipDataAnaysisMethodandRQandSQ} shows the RQs that are supposed to be answered using the extracted data. 

After data extraction, the data analysis was performed wherein we used open coding \& constant comparison methods \cite{seaman1999qualitative} and descriptive statistics \cite{easterbrook2008selecting} to analyze the extracted data items (see Table \ref{RelationshipDataAnaysisMethodandRQandSQ}) and answer the first three RQs, i.e., RQ1, RQ2, and RQ3 (see Section \ref{Introduction}).
To effectively code and categorize data, we used the qualitative data analysis tool MAXQDA. MAXQDA assists human annotators in labeling text segments within their contexts and assigning them to categories. Note that before the formal data analysis (manual coding), to reach an agreement about the data items that we extracted from the selected commits and issues, we first performed a pilot data analysis. This analysis process involves the following steps: (1) The first author checked and read a random sample of 10 commits and 10 issues. (2) The first author coded the extracted data with codes that succinctly summarize the data items (see Table \ref{RelationshipDataAnaysisMethodandRQandSQ}) for answering RQs. (3) The first author grouped all the codes into higher-level concepts and turned them into categories. The grouping process was iterative, in which the first author continuously went back and forth between the code, concepts, categories, and extracted data items to revise and refine the code, concepts, and categories. In order to improve the reliability of the results of the pilot data analysis, the first two authors held a meeting and followed a negotiated agreement approach~\cite{campbell2013coding} to compare the coding results, then discussed their disagreements and uncertain judgments about the data encoding results in an effort to reconcile them and arrive at a final version of the results of the pilot data analysis in which all discrepancies have been resolved. The first author carried on with the formal data analysis and followed the same steps used during the pilot data analysis. \textcolor{black}{To complement the qualitative analysis, we employed descriptive statistics~\cite{easterbrook2008selecting} to examine the frequency and distribution of results. Specifically, we used frequency counts and percentage distributions to summarize the occurrence of categories (e.g., categories of reused architectural solutions from SO in OSS projects) across our dataset (e.g., issues and commits)}. In the following paragraphs, we provide details of the formal data analysis process.

\textit{a) For analyzing RQ1 and RQ2}

As abovementioned, we utilized open coding \& constant comparison methods \cite{seaman1999qualitative} and descriptive statistics \cite{easterbrook2008selecting} to manually analyze the extracted data (i.e., the solved architectural problems for RQ1 and the \textcolor{black}{reused} architectural solutions for RQ2) as shown in Table \ref{RelationshipDataAnaysisMethodandRQandSQ}. With these RQs, we investigated issues and commits from two aspects, namely categorization of architectural problems encountered (RQ1) in these issues and commits and categorization of architectural solutions \textcolor{black}{reused} from SO to solve those issues and commits (RQ2). For example, regarding the categorization of architectural problems encountered, the first author referred to this issue\footnote{\url{https://github.com/SORMAS-Foundation/SORMAS-Project/issues/180}}: ``\textit{All communication between app and server is authenticated with the user's credentials (...) Use basic auth to secure transactions on the server side. Use basic auth to secure transactions on the app side. Use the provided user login to authenticate (...). Things to deal with when using basic auth: (1) In order to use the service, the client needs to keep the password somewhere in clear text to send it along with each request. (2) The verification of a password should be very slow (to counter brute force attacks), which would hamper the scalability of your service. On the other hand, security token validation can be quick. Source: \href{http://softwareengineering.stackexchange.com/questions/290511/\newline auth-options-for-distributed-systems/291030\#291030}{http://softwareengineering.stackexchange.com/questions/290511/\newline auth-options-for-distributed-systems/291030\#291030}}''. In this case, the first author summarized and encoded sentences as code ``using basic auth to secure the communication between app and server''. Subsequently, the first author grouped this code into a higher-level concept (i.e., ``authenticating communication between app and server''). Then, he applied constant comparison to compare the concepts identified in one summarized idea with the concepts that emerged from other summarized ideas to identify the concepts that have similar semantic meanings. He proceeded to group similar concepts into main categories. For example, the concept ``authenticating communication between app and server'' was merged into the category ``security issue'' (see Section \ref{ResultsOfRQ1}). As in the pilot data analysis, the first two authors held a meeting and followed a negotiated agreement approach~\cite{campbell2013coding} to compare the coding results, then discussed their disagreements and uncertain judgments about the data encoding results in an effort to reconcile them and arrive at a final version of the results of the formal data analysis for RQ1 in which all discrepancies had been resolved, therein improved the reliability of the analysis results for RQ1. The results of RQ1 and RQ2 are provided in Section \ref{ResultsOfRQ1} and Section \ref{ResultsOfRQ2}, respectively. \textcolor{black}{Moreover, to make the results of RQ1 and RQ2 more comprehensive and actionable, we further analyzed these two RQs to provide a mapping between the identified categories of architectural problems and the corresponding architectural solutions. Specifically, we examined the co-occurrence of problems and solutions (i.e., when certain solutions and problems appear together). 
The results and interpretations of this analysis are presented in Section \ref{Mapping_Problems_And_Solutions}.
}

\textcolor{black}{\textit{b) For analyzing RQ3}}

\textcolor{black}{This RQ investigates the ways of \textcolor{black}{reusing} architectural solutions from SO in OSS development. Similar to RQ1, the analysis employed open coding \& constant comparison methods \cite{seaman1999qualitative} and descriptive statistics \cite{easterbrook2008selecting}.  
The first author conducted a systematic analysis by following these steps: (i) \textit{Review of Commit, Issue, \textcolor{black}{and Posts}}: Commit messages, issue comments, \textcolor{black}{and SO posts} were examined to understand and gain insights into the rationale for adopting or implementing specific architectural solutions from SO. Examples of such solutions include introducing authentication, concurrency, load balancing, caching, and modifiability tactics, employing authentication protocols (e.g., OAuth 2.0), in-memory data stores (e.g., Redis, Memcached), dependency injection frameworks (e.g., Spring Framework), middleware (e.g., Node.js), APIs (e.g., REST API) to enhance quality attributes of systems such as security, scalability, extensibility, performance, and maintainability. (ii) \textit{Code Analysis}: Commit files, issue threads, \textcolor{black}{and SO answers} were examined to assess how architectural solutions from SO were integrated into project codebases. The analysis verified the implementation of these solutions, highlighting recurring patterns, methods, or practices, such as adapting solutions to align with project-specific contexts.}  

\textcolor{black}{For example, the following commit\footnote{\url{https://github.com/brave/brave-core/commit/e3b2ea79b67424484de5b09839ba8222fd261b40}} was analyzed: \textbf{Architectural Issue}: “\textit{New Tab Page WebUI: do not dispatch actions inside reducer before the reducer is finished. This is an anti-pattern and could introduce bugs. See discussion at \href{https://stackoverflow.com/questions/36730793/can-i-dispatch-an-action-in-reducer}{https://stackoverflow.com/questions/36730793/can-i-dispatch-an-action-in-reducer}.}”
Architectural Solution from SO: \textit{Starting another dispatch before your reducer is finished is an anti-pattern, because the state you received at the beginning of your reducer will not be the current application state anymore when your reducer finishes. But scheduling another dispatch from within a reducer is NOT an anti-pattern. In fact, that is what the Elm language does, and as you know \texttt{Redux} is an attempt to bring the Elm architecture to JavaScript. Here is a middleware (asyncDispatchMiddleware) that will add the property asyncDispatch to all of your actions (...)}. The SO post proposes a solution (an approach) that aligns with \texttt{Redux}’s architectural principles by using middleware to schedule asynchronous dispatches after the reducer finishes. This approach preserves the integrity of the reducer by avoiding side effects within reducer, ensuring state consistency.} 

\textcolor{black}{Upon reviewing this commit, the first author noted that the developers had translated the architectural solution description into code while reusing the ideas in the solution. Accordingly, these observations were encoded as “turning ideas in the solution into code”. This code was subsequently grouped into the category “Convert ideas in the architectural solutions into code” (see Section \ref{ResultsOfRQ3}). To reduce personal bias, the second author participated in validating the generated categories. Any disagreements were resolved through the negotiated agreement approach \cite{campbell2013coding}, ensuring the reliability of the analysis for RQ3, as with RQ1 and RQ2. The qualitative analysis of RQ3 identified five categories of ways that practitioners follow when \textcolor{black}{reusing} architectural solutions from SO in OSS development, as detailed in Section~\ref{ResultsOfRQ3}.}

\color{black}
\subsection{Survey Study}\label{Survey_Study}
\subsubsection{Collecting Data for the Survey}\label{CollectingSurveyData}
\color{black}
\textcolor{black}{As mentioned in Section \ref{Methodology}, we conducted an online questionnaire targeting practitioners contributing to OSS systems with two main objectives: (i) not to validate, but to complement the results of the mining study by eliciting additional insights directly from practitioners, such as information on architectural problems practitioners face in OSS development, and (ii) to answer the last two research questions (RQ4 and RQ5) of this study. Specifically, we designed and executed a descriptive survey \cite{personal2005} following the guidelines proposed by Kitchenham and Pfleeger \cite{personal2005}. The survey was conducted using Google Forms, which were made accessible online to participants. 
} 

\textbf{Structure of the survey questionnaire} \textcolor{black}{(see the survey questionnaire in Table \ref{surveyquestionnaire})}.\label{CreatingtheQuestionnaire} 
When creating our survey questionnaire, we ensured that the Survey Questions (SQs) were linked to our main RQs. Briefly, the survey starts with an Introduction page that explains the goal of the survey along with some preliminary definitions and the time frame (i.e., 3 to 5 minutes) that a participant could take to finish the survey. The survey comprises two sections: The first section includes Demographic Questions (DQs) (i.e., DQ1 to DQ5). For example, we asked about the role of the participants (e.g., developer, architect) and their years of experience in software development. The second section of the survey includes questions on \textcolor{black}{reusing} architectural solutions from SO (i.e., SQ7-SQ11). For example, Q7 asks about architectural problems that are solved by \textcolor{black}{reusing} architectural solutions from SO in OSS development, while SQ8 asks about types of \textcolor{black}{reused} architectural solutions from SO in OSS development. SQ9 asks about characteristics of architectural solutions from SO that practitioners consider when \textcolor{black}{reusing} these solutions in OSS development. SQ10 investigates how architectural solutions from SO are \textcolor{black}{reused} in OSS development, and SQ11 examines the challenges faced when \textcolor{black}{reusing} architectural solutions from SO in OSS development. In total, our survey comprises 12 questions, which are all mandatory except for two questions (i.e., SQ6 and SQ12). SQ6 asks for the participant's email address for sharing the survey results in case s/he wants them, and SQ12 asks for comments or remarks about relevant aspects (if any) on \textcolor{black}{reusing} architectural solutions from SO in software development that are possibly left uncovered.

Our survey includes a mix of closed-ended questions (i.e., SQ1, SQ2, SQ3, SQ4) and \textcolor{black}{semi-closed questions (i.e., SQ5, SQ7, SQ8, SQ10, SQ11)}. The closed-ended questions consist of a fixed set of multiple-choice items, numerical values, Yes/No answers, and Likert scale questions \cite{likert1932technique}. \textcolor{black}{For the semi-closed questions, we provided a randomly selected set of options based on, for example, categories identified in the mining study, along with an ``Other'' field to allow respondents to add their own opinions if none of the provided options applied. This approach aimed to minimize restrictions and potential bias in participants’ responses.} \textcolor{black}{Note that we chose not to share all categories from the mining study with the survey participants for two reasons: First, the survey was designed to complement the data mined from GitHub and enrich the findings of the mining study, with the goal of identifying new results, such as categories of architectural problems and challenges not captured in the mining study. Second, we anticipated that presenting all categories from the mining study to the survey participants would have hindered our ability to collect new insights from the survey participants.} 

\textbf{Target population recruitment}.\label{Targe population} 
The target population is the practitioners who \textcolor{black}{reused} and referenced SO posts (e.g., architectural solutions) in commits or issues mined from GitHub (see Section \ref{CollectingDataforMiningStudy}). Specifically, we used GitHub developers by collecting their email addresses provided in users’ profiles by checking the project repositories whose commits and issues were mined for our mining study (see Section \ref{CollectingDataforMiningStudy}). However, some users provided minimal information about their identity in their profile. This prevented us from identifying all potential practitioners. We further searched for those users on social media sites, such as LinkedIn, in order to validate their identity and gather missing background and contact information. After inspecting all user profiles, we found 1,015 valid user profiles, from which we could discern their identity and contact information. We note that the anonymity of these developers is preserved in both the manuscript and the accompanying dataset~\cite{dataset}. 

\textcolor{black}{To meaningfully engage practitioners, we designed a single Google Forms questionnaire containing identical questions for all participants. However, during recruitment, we sent personalized email invitations tailored to each recipient. These emails included the URL of a specific issue or commit in which the practitioner had been involved (e.g., as a reporter, assignee, or reviewer). This approach allowed participants to ground their responses in a familiar and relevant context, thereby enhancing both recall and response quality. Importantly, the customization was limited to the email content, not the questionnaire itself. This decision was motivated by two key factors: first, to prevent misunderstandings about which commit, issue, or architectural solution(s) the questionnaire referred to; and second, to align with established practices on customized survey invitations in prior research (e.g., \cite{martinez2021did, li2021understanding}). All responses were collected through a single Google Forms instance, ensuring uniformity, ease of data aggregation, and facilitating later data extraction and analysis via MS Excel.}

\textbf{Validation and evaluation of the survey design}.\label{EvaluatingandValidatingtheQuestionnaire} 
Before hosting the survey online for data collection, all authors thoroughly reviewed the survey design internally by cross-checking several parts of the survey protocol, such as survey objectives, survey questions, and RQs. Moreover, we also performed a pilot survey by sending invitations to ten participants through their email addresses that we gathered from GitHub and seven participants provided their responses to our pilot survey. Among the seven participants, five were professional developers and two were architects. During the pilot survey study, we examined various factors that could influence the survey outcomes, such as the comprehensibility of the survey questions and the time frame needed to answer the questions. Based on the pilot survey study responses, we refined a few survey questions to make them clearer and understandable.
 
\textbf{Filtering valid responses from the formal survey}.\label{ValidSurveyResponses} 
After sending the invitations through the gathered emails, we obtained 229 responses. Among those, we excluded the inconsistent, randomly filled, or meaningless responses (i.e., 2 responses). We finally got 227 valid responses. \textcolor{black}{Note that the number of practitioners (i.e., 1,015) who received the survey questions is greater than the number of issues and commits analyzed (984 commits and issues). This is because, during the recruitment phase, we reached out to multiple practitioners (e.g., developers) associated with each issue or commit, such as the reporter (the individual who initially created or reported the issue), assignee (the developer responsible for resolving the issue), reviewer (who reviews code changes linked to issues via commits), and maintainer (who may triage issues, assign tasks, enforce contribution guidelines, and make final decisions), rather than limiting contact to a single individual per data point. Our intention was to maximize the chances of receiving responses, especially in cases where one project member might miss or choose not to respond to the survey invitation.}

\small 
\begin{table}
\color{black}
\caption{\textcolor{black}{Survey questionnaire}}
\label{surveyquestionnaire}
\begin{tabular}{|p{8.7cm}|}
\hline
\textbf{Demographic} \\
\hline
\textbf{Q1}. Which country are you working in? \\
\hline
\textbf{Q2}. What is your main role in open source software development? \\
\(\circ\) Project manager 
\(\circ\) Team Leader
\(\circ\) Requirements Engineer 
\(\circ\) Architect
\(\circ\) Developer
\(\circ\) Consultant 
\(\circ\) Tester
\(\circ\) Other \_\_\_\_\_\_\\ 
\hline
\textbf{Q3}. How many years have you been involved in open source software development? \\
\(\circ\) 0 years $\leq$ 1 \(\circ\) 1 years $\leq$ 2 \(\circ\) 2 years $\leq$ 4 
\(\circ\) 4 years $\leq$10 \\
\(\circ\) 10 years $\leq$ 20 \(\circ\) years $>$ 20 \\ 
\hline
\textbf{Q4}. What is the size of the largest open source software project you participated in (number of people involved)? \\
\(\circ\) $< 5$ \(\circ\) 5-10 \(\circ\) 10-50 \(\circ\) 50-100 \(\circ\) $>$ 100 \\ 
\hline
\textbf{Q5}. What are the domains or areas of your open source software projects? (select all that apply) \\
\(\circ\) Consulting and IT service \(\circ\) Embedded system \(\circ\) E-commerce\\ 
\(\circ\) Financial \(\circ\) Healthcare \(\circ\) Retail \(\circ\) Insurance \(\circ\) Telecommunication \\
\(\circ\) Software framework \(\circ\) Software library\\
\(\circ\) Other \_\_\_\_\_\_\\ 
\hline
\textbf{Q6}. \textbf{Optional}: Do you want to get the results of this survey? If so, please provide your email address. (Email will be kept confidential) \\
\hline
\textbf{Architectural Solutions Utilization Related Questions} \\
\hline
\textbf{Q7}. What are the architectural problems that you solve by utilizing architectural solutions from SO and SWESE in open source software development? (select all that apply) \\
\(\Box\) Security issue \(\Box\) Performance issue \(\Box\) Scalability issue \\
\(\Box\) Reliability issue \(\Box\) Component design issue \\
\(\Box\) Component communication issue \\
\(\Box\) Architecture deployment issue \(\Box\) Architecture pattern design issue \(\Box\) Architecture anti-pattern issue
\(\Box\) Architecture documentation issue \(\Box\) Incompatibility issue \\
\(\Box\) Other \_\_\_\_\_\_\\ 
\hline
\textbf{Q8}. What are the architectural solutions that you utilize from SO and SWESE in open source software development? (select all that apply) \\
\(\Box\) Architectural patterns (e.g., Model-View-Controller pattern) \\
\(\Box\) Architectural tactics (e.g., heartbeat) \(\Box\) Architectural refactoring \\
\(\Box\) Use of libraries (e.g., React) \(\Box\) Use of APIs (e.g., RESTful) \\
\(\Box\) Use of frameworks (e.g., Django) \\
\(\Box\) Use of protocols (e.g., OAuth 2.0)\\ 
\(\Box\) Other \_\_\_\_\_\_\\ 
\hline
\textbf{Q9}. What characteristics do you want while utilizing architectural solutions from SO and SWESE in open source software development? (select all that apply) \\
\(\Box\) If architectural solution description includes a sample of architecturally-relevant code \\
\(\Box\) If architectural solution description includes one or more architectural diagrams (e.g., component diagram) \\
\(\Box\) If architectural solution description provides the design context \\
\(\Box\) If architectural solution description includes a concise summarization \\
\(\Box\) If architectural solution description provides the source of the solution \\
\(\Box\) If architectural solution is accepted by the seeker of the question \\ 
\(\Box\) Other \_\_\_\_\_\_\\ 
\hline
\textbf{Q10}. What are the ways you utilize architectural solutions from SO and SWESE in open source software? (select all that apply) \\
\(\Box\) Adapt the architectural solutions to fit in the project context \\
\(\Box\) Use the architectural solutions directly in the project \\
\(\Box\) Convert ideas in the architectural solutions into detail design \\
\(\Box\) Convert ideas in the architectural solutions into code\\
\(\Box\) Learn the architectural knowledge in the architectural solutions\\
\(\Box\) Other \_\_\_\_\_\_\\ 
\hline
\textbf{Q11}. What are the challenges you face when utilizing architectural solutions from SO and SWESE in open source software development? (select all that apply) \\
\(\Box\) Architectural solutions are not provided in the design context that I need \\
\(\Box\) It takes too much time to adapt the architectural solution to address my design concerns \\
\(\Box\) Hard to figure out the up-to-dateness (e.g., lastest technologies) of the architectural solution \\
\(\Box\) Hard to judge the credibility of the architectural solution \\
\(\Box\) Contradictions (e.g., both positive and negative impact to certain quality attribute) in the architectural solution description\\ 
\(\Box\) Other \_\_\_\_\_\_\\ 
\hline
\textbf{Q12}. \textbf{Optional}: Do you have any further comments about utilization of architectural solutions from SO and SWESE in development? \\
\hline
\end{tabular}
\end{table}\color{black}
\normalsize 

\begin{table}
\footnotesize
\caption{Relationship between data analysis methods, research questions, and survey questions}
\label{RelationshipDataAnaysisMethodandRQandSQ}
\begin{tabular}{m{0.1cm}<{\centering}m{2cm}m{2cm}m{0.25cm}m{1.13cm}m{0.9cm}<{\centering}}
\hline
\textbf{\#}  & \textbf{Data item}         & \textbf{Data analysis approach}   &  \textbf{RQ}  & \textbf{Survey question} & \textbf{\textcolor{black}{Data source}}\\ \midrule
D0 & Demographics                         & Descriptive statistics                                        &      & DQ1-DQ5 & \textcolor{black}{Survey}  \\\hline
D1 & The solved architectural problem     & Open coding \& constant comparison and Descriptive statistics & RQ1  & SQ7     & \textcolor{black}{GitHub \& Survey} \\\hline
D2 & The \textcolor{black}{reused} architectural solution  & Open coding \& constant comparison and Descriptive statistics & RQ2  & SQ8     & \textcolor{black}{GitHub \& Survey} \\\hline
D3 & The way of \textcolor{black}{reusing} architectural solution             & Open coding \& constant comparison and Descriptive statistics  & RQ3 & SQ10 & \textcolor{black}{GitHub \& Survey} \\\hline
D4 & The characteristic of architectural solution  & Open coding \& constant comparison and Descriptive statistics  & RQ4 & SQ9 & \textcolor{black}{Survey}\\\hline
D5 & The challenge in \textcolor{black}{reusing} architectural solution     & Open coding \& constant comparison and Descriptive statistics  & RQ5 & SQ11 & \textcolor{black}{Survey} \\\hline
\end{tabular}
\end{table}

\color{black}
\subsubsection{Survey Data Extraction and Analysis}\label{Data_extraction_analysis_for_survey}
\color{black}
We extracted the survey data based on the data items listed in Table \ref{RelationshipDataAnaysisMethodandRQandSQ} and analyzed the responses using two approaches (see Table \ref{RelationshipDataAnaysisMethodandRQandSQ}). For open text fields (“Other”), we employed open coding and constant comparison, following the same procedures used in the mining study (e.g., encoding extracted data items and grouping similar codes into high-level concepts and categories) to analyze data corresponding to each RQ. Descriptive statistics \cite{wohlin2003empirical} were employed to better understand the frequency and distribution of responses. The following paragraph provides details on survey data analysis.

As mentioned earlier, the survey participants were provided with several candidate answers (i.e., categories derived from the mining study) to choose from when responding to survey questions. Additionally, an open text field (“Other”) was provided for each question, allowing participants to offer their own relevant answers. In the open text field, we collected 41 answers for architectural problems, 28 answers for architectural solutions, 12 answers for ways of \textcolor{black}{reusing} architectural solutions, 19 answers for characteristics of architectural solutions, and 17 answers for challenges in architectural solutions utilization for RQ1, RQ2, RQ3, RQ4, and RQ5, respectively. \textcolor{black}{The analysis and mapping of participant responses to categories from the mining study involved the following steps: (i) Each survey response was independently analyzed by the first author to assess its alignment with the categories obtained from the mining study or determine whether it introduced a new category using open coding technique. (ii) To enhance the reliability of the survey data analysis, the first two authors employed a negotiated agreement approach \cite{campbell2013coding}, meeting to compare coding results, resolve discrepancies, and finalize a reconciled set of analysis results. (iii) The finalized categories from the survey data analysis were then compared with the categories derived from the mining study using constant comparison technique. This process identified overlapping categories for merging and differing categories for introducing new ones. (iv) This involved recording how ambiguous or unusual cases were handled and providing clear explanations for creating new categories when needed.}

\textcolor{black}{During the survey data analysis, we observed that most of the answers from the open text field fell or could be mapped into} the categories identified in the mining study. For example, when analyzing the responses for RQ1, one developer wrote: \faHandORight \hspace{0.5mm} “\textit{Most of our applications were developed using legacy techniques several years ago, and some of them did not follow robust design patterns described in well-known software engineering books (...) we need to maintain them and improve their qualities (...). I got interested in the StackOverflow forum after I saw some StackOverflow solutions and snippets in several projects in GitHub and software documentation, so I look into it a bit more and used some of the workarounds that I saw}” (P225). The first author summarized and coded this response as “maintaining application to improve quality”, then grouped it under the higher-level concept “maintaining application” Using constant comparison, similar concepts were identified and merged into the “maintainability issue” category (see Table \ref{CategoriesOfArchitecturalProblems}). 

Among the survey responses obtained from the open-text field, we identified one new category of architectural problems, no new categories of architectural solutions, and no new potential ways of \textcolor{black}{reusing} architectural solutions that were not part of the categories identified in our mining study. Additionally, the qualitative analysis of the survey responses revealed nine characteristics of architectural solutions from SO and seven categories of challenges developers face when \textcolor{black}{reusing} these solutions in OSS development. 
\textcolor{black}{In Section \ref{Results}, we provide details on how the results of the survey study were integrated with the results of the mining study. The valid survey responses, along with encoded data, are provided in our dataset as an MS Excel file~\cite{dataset}.}

\section {Results}\label{Results}
\textcolor{black}{This section presents the results of our mixed-methods study, including the mining study (Section~\ref{ResultsOfRQ1_to_RQ4}) and the survey study (Section~\ref{ResultsOfRQ4_to_RQ5}), organized by the five RQs introduced in Section~\ref{Introduction}. We begin by explaining how the two methods complement each other in addressing these RQs.}

As stated in Section~\ref{Methodology}, the first three RQs (RQ1, RQ2, and RQ3) are addressed using data from GitHub commits and issues, as well as the survey responses from OSS practitioners. The last two RQs (RQ4 and RQ5) are particularly answered through the survey data. 

\textcolor{black}{The results from both methods provide complementary insights. The mining study identified recurring patterns in architectural problems, solutions, and ways of \textcolor{black}{reusing} architectural solutions from SO across a broad set of projects. Meanwhile, the survey study added depth and context by capturing practitioners' experiences and perceptions. Specifically, the two methods support each other by offering complementary insights that neither could achieve alone. The survey responses revealed one new category of architectural problems not captured in the mining study, as well as nine categories of characteristics of architectural solutions and their descriptions developers consider important for reusing those solutions in OSS development and seven categories of challenges OSS developers face when \textcolor{black}{reusing} these solutions. No new categories of architectural solutions or ways of \textcolor{black}{reusing} solutions emerged from the survey. These new categories identified in the survey study complement the findings from the mining study and offer a more nuanced understanding of the \textcolor{black}{reuse} of architectural solutions from Stack Overflow in OSS development, thereby enhancing the comprehensiveness of our answers to the research questions.}

\textbf{Demographics of Survey Participants}. We first summarize the demographic information of the survey participants in the following.

\begin{figure} [h]
  \centering
  \includegraphics[width=1\linewidth]{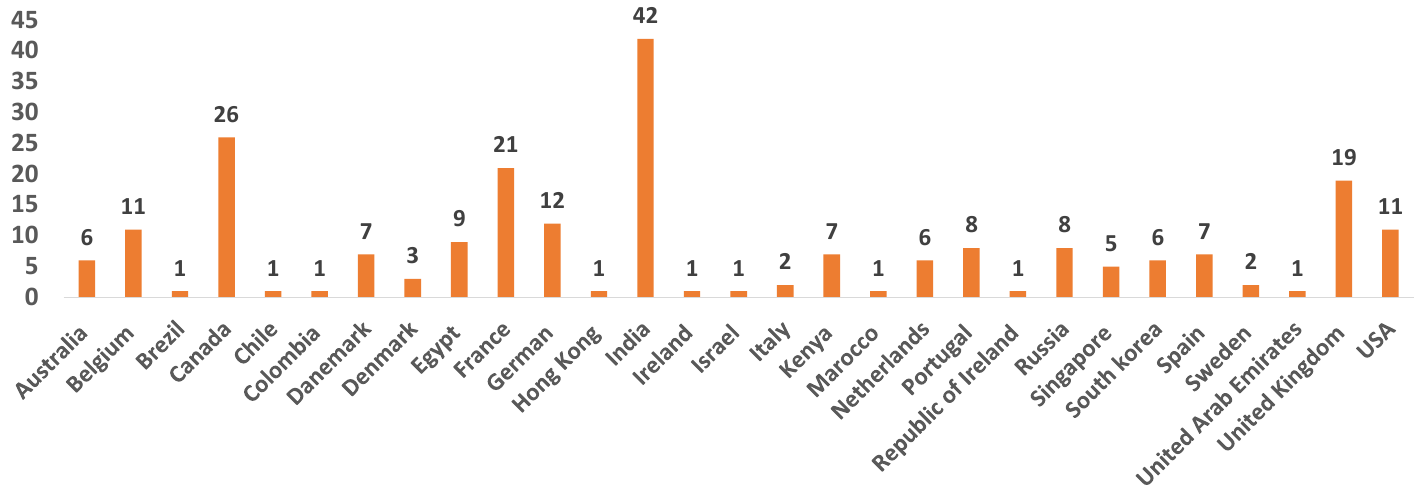}
      \caption{Geographic distribution of the survey participants}
  \label{GeographicDistribution}
\end{figure}

\textit{Country}: The survey participants come from 6 continents (37 countries). Most of the participants come from India (18.5\%, 42 out of 227), followed by Canada (11.5\%, 26 out of 227) (see Figure \ref{GeographicDistribution}). 

\begin{figure} [h]
  \centering
  \includegraphics[width=1\linewidth]{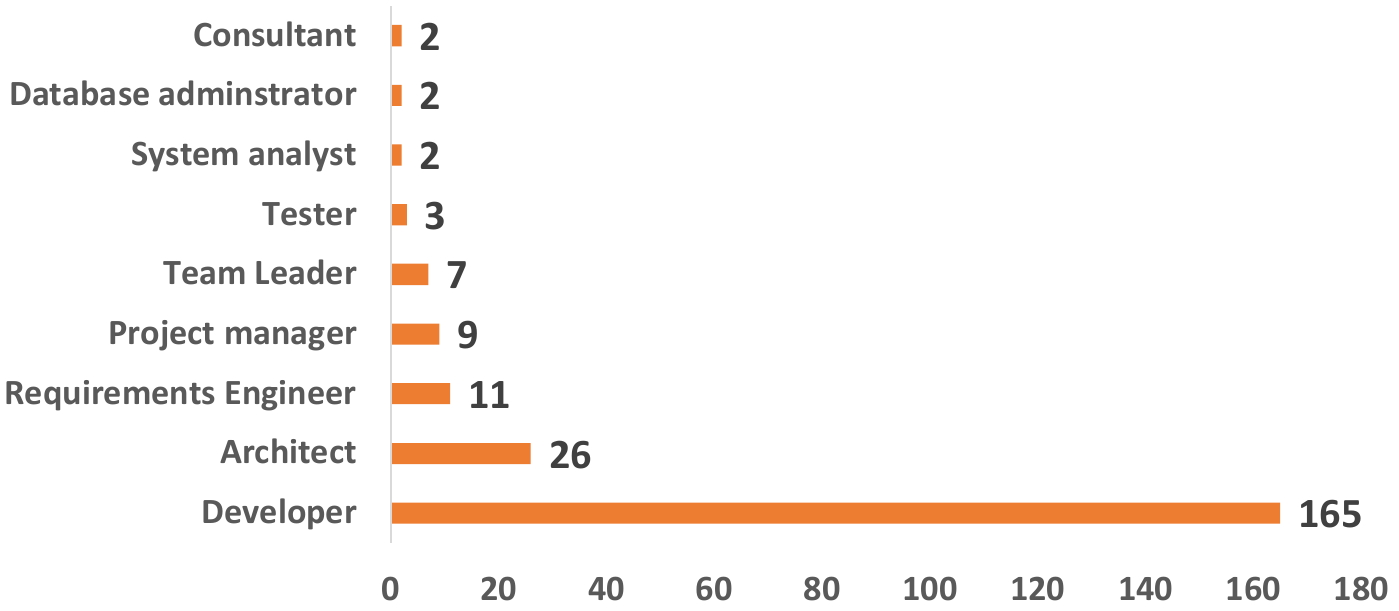}
      \caption{\textcolor{black}{Roles of the survey participants}}
  \label{RoleAndOrganizationDomain}
\end{figure}

\textit{Role}: As shown in Figure \ref{RoleAndOrganizationDomain}, the survey participants cover different roles, such as developer, architect, requirements engineer, project manager, and team leader. A large number of participants worked as software developers (72.7\%, 165 out of 227), followed by architects (11.5\%, 26 out of 227) and requirements engineers (4.9\%, 11 out of 227). 

\textit{Experience}: The distribution of experience of the participants is: 2 to 4 years (11 out of 227, 4.8\%), 4 to 10 years (82 out of 227, 36.1\%), 10 to 20 years (93 out of 227, 40.9\%), 20+ years (41 out of 227, 18.1\%). Most of the participants have more than 10 years of experience working in software development, i.e., 10 to 20 years (93 out of 227, 40.9\%).  

\textit{Project size}: The participants reported different sizes of the largest projects they had worked on (number of people involved). Most of the participants, i.e., 46.3\% (105 out of 227), worked on projects where people ranged between 10-50, followed by 22.5\% (51 out of 227) who worked on projects where people ranged between 50-100. 

\begin{figure} [h]
  \centering
  \includegraphics[width=1\linewidth]{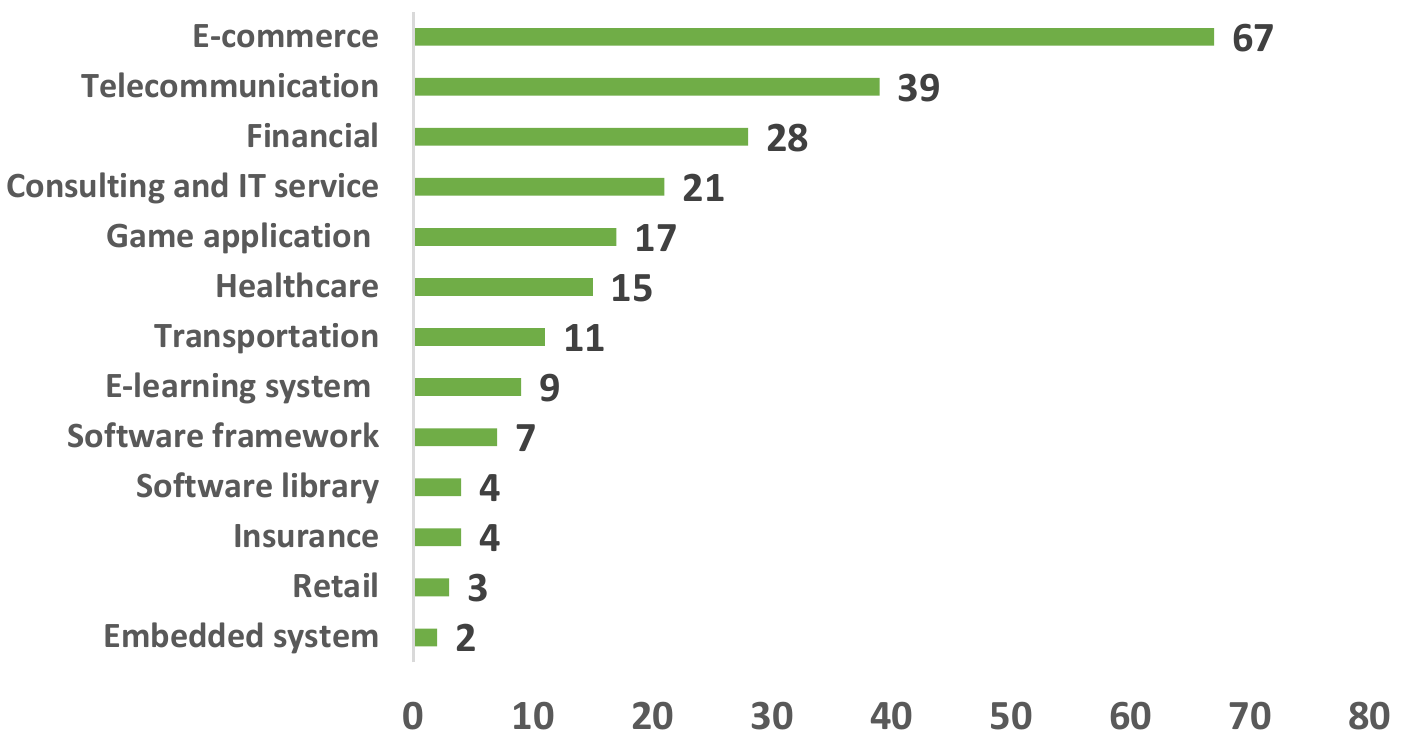}
      \caption{\textcolor{black}{Application domains the participants worked on}}
  \label{OrganizationAndApplicationDomain}
\end{figure}

\textit{Application domain}: The domains of the applications that the participants worked on vary significantly, including E-commerce, telecommunication, embedded systems, financial, healthcare, retail, insurance, and game systems (see Figure~\ref{OrganizationAndApplicationDomain}. A diversified population of participants gives us confidence that our recruited participants are representative of the survey target population.

\subsection{RQs Answered by the Mining and Survey Studies}\label{ResultsOfRQ1_to_RQ4}
\subsubsection{Architectural Problems Solved Through the \textcolor{black}{Reuse} of Architectural Solutions from SO (RQ1)}\label{ResultsOfRQ1}
Through our mining study, we identified 14 categories of architectural problems encountered by OSS practitioners that were addressed through the reuse of architectural solutions from SO. Additionally, with the survey responses, we got one additional category of architectural problem, namely \textit{Accessibility issue} that practitioners solved by \textcolor{black}{reusing} architectural solutions from SO. Table \ref{CategoriesOfArchitecturalProblems} presents these architectural problems addressed by architectural solutions from SO, as identified through the mining and survey studies. \textcolor{black}{Note that a single commit or issue may report multiple architectural problems that were solved by \textcolor{black}{reusing} architectural solutions from SO. We counted each architectural problem reported in a commit or issue separately, which is why the total frequency in Table \ref{CategoriesOfArchitecturalProblems} exceeds the number of commits and issues (984) analyzed in our study.} Moreover, we observed that some participants mentioned more than one architectural problem they encountered and solved by \textcolor{black}{reusing} architectural solutions from SO. As a result, the total frequency in the survey study exceeds the number of survey participants (227) as shown in Table \ref{CategoriesOfArchitecturalProblems}. 
In the following paragraphs, we provide a detailed description of the categories of architectural problems solved by architectural solutions from SO. To clearly present these categories and highlight the insights, we use the following notations: \textit{[MS]} \faGithub \hspace{0.5mm} denotes an example of a GitHub issue or commit that fits into a category of architectural problem, while \textit{[S]} \faHandORight \hspace{0.5mm} represents the survey participant's experience or opinions related to the architectural problem.

\textit{\textbf{Component design issue}}. A software system may be designed with various components (or modules) to carry out functions and tasks of the system \cite{SA2012}. During the software architectural design, different components are identified and designed to satisfy the key functionalities of the system \cite{SA2012}. However, designing these components to meet both functional and non-functional requirements within the application stack can be challenging for developers \cite{karthik2019automatic}. As shown in Table \ref{CategoriesOfArchitecturalProblems}, \textit{Component design issue} is the most common category of architectural problems that OSS developers face and solve by \textcolor{black}{reusing} architectural solutions from SO. We identified several component design issues, including component does not return components in the correct order, components re-rendering issues, components return warnings, ModuleNotFoundError, components return incorrect element errors, and component loading issues. 

\textit{[MS]}: \faGithub \hspace{0.5mm} In this commit\footnote{\url{https://github.com/codenameone/CodenameOne/commit/2ef6a77e4f8ac3475a8c05541b8ac829099ac9a9}}, developers of the \textit{CodenameOne} project \textcolor{black}{reused} an architectural solution from SO to solve an issue in which a component was not returning components in the correct order (i.e., in the same order in which they are shown in the given Container): ``\textit{Fixed an issue with ComponentSelector. Specifically: \textit{ComponentSelector} doesn't return \textit{Components} in the correct order discussed here: \href{https://stackoverflow.com/questions/57734646/componentselector-doesnt-return-components-in-the-correct-order}{https://stackoverflow.com/questions/57734646/componentselector-doesnt-return-components-in-the-correct-order}}.''

\textit{[S]} \faHandORight \hspace{0.5mm} ``\textit{(...) we often face poor component design problems, which may grow as the application size grows. We recognize that the poor designs of components may lead to mysterious problems in the system (...)}.'' P197  

\textit{\textbf{Architectural anti-pattern}}. An architectural anti-pattern arises from decisions that, either consciously or unconsciously, undermine critical system-wide quality attributes such as maintainability, evolvability, and security \cite{li2014architectural}. These anti-patterns often result from violations of established architectural principles or rules \cite{verdecchia2020architectural}. The architectural problems in this category are mainly related to cyclic component dependency issue \cite{baldwin2000design}, modularity violation \cite{baldwin2000design}, circular (or cyclic) module import problem, unstable inheritance hierarchy \cite{liskov1987keynote}, duplicate class/component issue, scattered functionalities \cite{garcia2009toward}, and component overload \cite{macia2012automatically}.

\textit{[MS]}: \faGithub \hspace{0.5mm} In this commit\footnote{\url{https://github.com/streamr-dev/core-api/commit/ee0f2f63aaa1c35008d8e94ea1064e4cf456a6b9}}, developers of the \textit{Streamr Engine and Editor} project \textcolor{black}{reused} an architectural solution from SO to solved cyclic component/service dependency issue: ``\textit{Fetching a Canvas from database can be very expensive if its serialization is
large (e.g tens of megabytes.) This is especially problematic when fetching lists of Canvases. Unfortunately Grails only offers lazy fetching for associations and not within a domain class/table. So I made a separate Serialization domain class that is associated with (and dependent on) a Canvas. Not the best possible solution to: (...) * StreamService: remove circular dependency, read more: \href{http://stackoverflow.com/questions/8002951/grails-service-class-cross-ref}{http://stackoverflow.com/questions/8002951/grails-service-class-cross-ref}}.'' 

\textit{[S]} \faHandORight \hspace{0.5mm} ``\textit{(...) a very naive design would have problems like circular dependencies (between modules or components, classes, whatever). I liked and employed some interesting workarounds discussed on this community forum (Stackoverflow) to tackle such glitches in our projects (...)}.'' P156  

\textit{\textbf{Security issue}} refers to vulnerabilities or flaws in the system's architecture that expose it to threats compromising critical quality attributes like confidentiality, integrity, and availability. These issues often originate from insufficient planning for security during the design phase, where the focus on functional requirements may overshadow security consideration. We collected various security problems that OSS developers solved using architectural solutions from SO, such as issues in enabling authentication mechanisms, authorization mechanisms, encryption and decryption mechanisms, and issues in enabling digital certificates (such as SSL and TSL). 

\textit{[MS]}: \faGithub \hspace{0.5mm} In this issue\footnote{\url{https://github.com/SORMAS-Foundation/SORMAS-Project/issues/180}}, the developers of the \textit{SORMAS} project \textcolor{black}{reused} an architectural solution from SO to solve the authentication issue: ``\textit{All communication between app and server is authenticated with the user's credentials [3] \#180 (...) Things to deal with when using basic auth: In order to use the service, the client needs to keep the password somewhere in clear text to send it along with each request. The verification of a password should be very slow (to counter brute force attacks), which would hamper scalability of your service. On the other hand, security token validation can be quick source: 
\href{https://softwareengineering.stackexchange.com/q/290511/422568}{https://softwareengineering.stackexchange.com/q/290511/422568}}.'' 

\textit{[S]} \faHandORight \hspace{0.5mm} ``\textit{We have to keep up with the various security patches (...) in my understanding, providing security to complex systems is the biggest challenge that we come across most of the time}.'' P112  

\textit{\textbf{Component communication issue}}. Software applications consist of diverse components, including software and hardware elements, that must interact and communicate effectively to address various design concerns \cite{karthik2019automatic}. Communication between components typically occurs via protocols (e.g., HTTP, AMQP, WebSockets, TCP), frameworks (e.g., gRPC), or APIs (e.g., REST), chosen based on the specific characteristics and requirements of the components involved. We collected various component communication issues, such as HTTP inter-component communication error, grpc inter-component communication error, TCP inter-component communication issue, and WebSocket error in inter-component communication, that OSS developers face and solved by \textcolor{black}{reusing} architectural solutions from SO. 

\textit{[MS]}: \faGithub \hspace{0.5mm} In this issue\footnote{\url{https://github.com/TNO/knowledge-engine/issues/449}}, developers of the \textit{Knowledge Engine} project \textcolor{black}{reused} an architectural solution from SO to solve HTTP inter-component communication error: ``\textit{In distributive mode, a ‘HTTP\/1.1 header parser received no bytes' error occurs during \textit{inter-KER communication} (...) Why this occurs and the solution is explained in this SO question: \href{https://stackoverflow.com/questions/74215117/flickering-httpclient-sometimes-throwing-ioexception/74233477\#74233477}{https://stackoverflow.com/questions/74215117/flickering-httpclient-sometimes-throwing-ioexception/74233477\#74233477}}.'' 

\textcolor{black}{\textit{[S]} \faHandORight \hspace{0.5mm} ``\textit{The one I usually faced, is designing react components and their communication, such as passing one react component into another react component to transclude the first component's content}.'' P132} 

\textit{\textbf{Performance issue}}. This category includes problems related to poor performance of some components (e.g., database servers) of the system or the system as a whole due to, for example, weak or bad design techniques used for addressing the performance requirement.

\textit{[MS]}: \faGithub \hspace{0.5mm} In this commit\footnote{\url{https://github.com/diegorcampos/obs-studio./commit/0e57a7beef8e7fa530f12fcf6dd3c3f1de9a5094}}, developers of the \textit{OBS Studio} project \textcolor{black}{reused} architectural solution from SO to resolve the performance issue: ``\textit{UI: Fix performance issues with the Log Viewer. This commit includes two big changes (...). Fix 1: significantly improves real-time performance (...), to the point that the UI no longer freezes if the viewer is open and the log is being spammed. It also improves initial launch speed (...). Reference: \href{https://stackoverflow.com/a/54501760/2763321}{https://stackoverflow.com/a/54501760/2763321}. Fix 2: completely eliminates delay when opening the viewer (...) Reference: \href{https://stackoverflow.com/a/17466240/2763321}{https://stackoverflow.com/a/17466240/2763321}}.''  

\textcolor{black}{\textit{[S]} \faHandORight \hspace{0.5mm} ``\textit{I use design solutions from Stack Overflow for optimizing my apps in terms of size and performance, especially Single Page JavaScript apps that need to be fast and lightweight}.'' P115}

\textit{\textbf{Architectural pattern design issue}}. The issues in this category are related to poor architecture pattern design. 

\textit{[MS]}: \faGithub \hspace{0.5mm} In this commit\footnote{\url{https://github.com/raynathanlow/sort-plus/commit/1a7c748dda77540e007c8843a2361de9b94c3490}}, developers of sort-plus project referred to an architectural solution from SO to solve the Client-Server design pattern issue:  ``\textit{Client and server shouldn't use the same route because it affects how the browser will know what data to get from its cache (...)  \href{https://stackoverflow.com/a/56022718}{https://stackoverflow.com/a/56022718}}.'' 

\color{black}
\textcolor{black}{\textit{[S]} \faHandORight \hspace{0.5mm} ``\textit{Ops, SO is my all time savior. My job is to design and structure desktop apps with, such as WPF which is built on the MVVM design pattern. I also solve several issues in design pattern like this for Windows Forms}}.'' P95

\textcolor{black}{\textit{\textbf{Architectural deployment issue}} arises when deployment-phase architectural decisions compromise functionality, performance, scalability, or maintainability. Such challenges often result from misconfigurations, component incompatibilities, scaling efforts, or infrastructure changes, particularly when the testing environment fails to replicate production, disrupting live component interactions}. 

\textcolor{black}{\textit{[S]} \faHandORight \hspace{0.5mm} ``\textit{I had a couple of things to do: - \textit{Deploy a full-stack app with `client/` `server/` directory structure/architecture} \& two packages. jsons to Heroku and Build TypeScript in a Heroku project. I struggled and mercifully had to stumble with this solution: Use `Heroku-postbuild` from a top-level package.json and reference the nested package.jsons in it: \href{https://stackoverflow.com/questions/47124869/deploying-a-full-stack-node-app-npm-package-json-architecture}{https://stackoverflow.com/questions/47124869/deploying-a-full-stack-node-app-npm-package-json-architecture})}.'' P147}

\textcolor{black}{\textit{\textbf{Architectural documentation issue}} occurs when architectural design is poorly recorded, described, or shared due to factors such as missing or outdated documentation, undocumented dependencies between components, unclear or inconsistent documentation practices, or the absence of standardized methods for documenting architecture. 
} 

\textit{[MS]}: \faGithub \hspace{0.5mm} In this commit\footnote{\url{https://github.com/longxinaghua/yii2/commit/4afd9f112b0a642d4aff6f8befdc9438e2f1e811}}, developers of the \textit{Yii 2} project followed an architectural solution from SO\footnote{\url{https://stackoverflow.com/a/5864000/12381813}} to update the architectural documentation: ``\textit{Update \textit{structure-controllers.md: In MVC the ``M'' is a model\-layer\-, and not just a collection of models}. If we don't teach this to new users, they will quickly end up with heavily bloated model classes. Maybe this deserves some attention in other introduction pages as well. \href{http://stackoverflow.com/questions/5863870/how-should-a-model-be-structured-in-mvc}{http://stackoverflow.com/questions/5863870/how-should-a-model-be-structured-in-mvc}.''}

\textcolor{black}{\textit{[S]} \faHandORight \hspace{0.5mm} ``\textit{I need to document several components of our systems. Not doubt, I google at SO}.'' P68}

\textcolor{black}{\textit{\textbf{Incompatibility issue}} refers to problems that occur when two or more subsystems, programs, or technologies fail to function as expected while operating on the same platform or within the same environment, such as hardware, software, or operating systems.} 

\textit{[MS]}: \faGithub \hspace{0.5mm} In this issue\footnote{\url{https://github.com/flutter/flutter/issues/152150}}, developers \textcolor{black}{reused} an architectural solution from SO to resolve an architecture incompatibility in the Flutter package for iOS, specifically with the arm64 architecture in the development of ``\textit{flutter}'' project: ``\textit{[IOS]How to make my own flutter package support ios arm64 architecture? \#152150 (...) 
\href{https://stackoverflow.com/questions/77139621/xcode-15-undefined-symbols-linker-command-failed-with-exit-code-1-use-v-to-s}{https://stackoverflow.com/questions/77139621/xcode-15-undefined-symbols-linker-command-failed-with-exit-code-1-use-v-to-s}. This method solved it (...).}'' 

\textcolor{black}{\textit{[S]} \faHandORight \hspace{0.5mm} ``\textit{Incompatibility of the solution with the existing system elements (e.g., components)}.'' P96}

\textcolor{black}{\textit{\textbf{Reliability issue}} occurs when a software system or its components fail to consistently perform their intended functions without errors over time. Such issues are critical in mission-critical systems, where downtime or errors can have severe consequences. Reliability issues can arise in various situations, such as when the system is unable to handle the expected load due to insufficient resources (e.g., memory, processing power, storage), resulting in system crashes.}

\textit{[MS]}: \faGithub \hspace{0.5mm} In this commit\footnote{\url{https://github.com/enkessler/childprocess/commit/82afeebca8a158c91bda6e320b03881df6599a49}}, the developers of the \textit{childprocess} project \textcolor{black}{reused} an architectural solution from SO for improving reliability at the system level, especially in the context of process management: ``\textit{Replace all backends by Process.spawn for \textit{reliability} (...). Try latest JRuby * Add CI workflow * Use Process.spawn, it is enough to implement ChildProcess (..) * Use `taskkill` to kill a process tree on Windows * Inspired by \href{https://stackoverflow.com/a/61113184/388803}{https://stackoverflow.com/a/61113184/388803}}.'' 

\textcolor{black}{\textit{\textbf{Extensibility issue}} occurs when a system or architecture cannot be easily modified to accommodate new features, functionalities, or requirements without significant rework. This often results from a lack of flexibility, modularity, or tightly coupled components.} 

\textit{[MS]}: \faGithub \hspace{0.5mm} In this commit\footnote{\url{https://github.com/Mystfit/Showtime-Cpp/commit/23642988e23c87b8530e86fa2896435e7e57e425}}, the developers of the \textit{Showtime\-Cpp} project \textcolor{black}{reused} an architectural solution from SO, incorporating mixins and callbacks as a design pattern, to make the system more modular, reactive, and adaptable to changes. This approach influences how the system will scale, be extended, and interact with other components in the future: ``\textit{Starting a mixin for ZstPlugs that will make it an extensible reacting plug with callbacks. \href{https://stackoverflow.com/questions/18773367/what-are-mixins-as-a-concept}{https://stackoverflow.com/questions/18773367/what-are-mixins-as-a-concept}}.''  

\textit{[S]} \faHandORight \hspace{0.5mm} ``\textit{(...) we always want to expand our applications to satisfy our customers' needs. I use the solutions from stack overflow while trying to add some components to our applications}.'' P45 

\textcolor{black}{\textit{\textbf{Scalability issue}} arises when a system or architecture cannot efficiently handle increased load, traffic, or user demands as it grows. This typically results from limitations in the system’s design, such as poor vertical or horizontal scaling strategies, unoptimized data handling, or specific bottlenecks in critical components like databases, network communication, or processing power. These bottlenecks restrict the system’s ability to scale effectively as the load increases.} 

\textit{[MS]}: \faGithub \hspace{0.5mm} In this commit\footnote{\url{https://github.com/avsm/eeww/commit/0008ace7a936477e0dfc040af9aa1d23af2dab5b}}, the developers of the \textit{Eeww} project \textcolor{black}{reused} an architectural solution from SO\footnote{\url{https://stackoverflow.com/a/14388707/12381813}} to build scalable network servers, which addresses a scalability concern in the architecture of network server systems. By enabling socket sharding across multiple threads or child processes, the solution helps distribute the load more efficiently and improve the scalability of the server: ``\textit{This commit adds the option of setting SO\_REUSEPORT socket option in linux\_uring backend (...) SO\_REUSEPORT can be used to build \textit{scalable} network servers in linux version 3.9 and over. This is so because reuse port is used to enable socket sharding on multiple threads or child processes. Listening sockets with SO\_REUSEPORT enabled are further load balanced by the kernel (...) Reference: \href{https://stackoverflow.com/questions/14388706/how-do-so-reuseaddr-and-so-reuseport-differ}{https://stackoverflow.com/questions/14388706/how-do-so-reuseaddr-and-so-reuseport-differ}}.'' 

\textit{[S]} \faHandORight \hspace{0.5mm} ``\textit{As a requirement engineer, I do search for solutions for scalability (...) design concerns. In addition, I do pick up the SO URL for later reference}.'' P58 

\textcolor{black}{\textit{\textbf{Maintainability issue}} arises when the system design lacks flexibility, modularity, or clear separation of concerns. For example, a system with tightly coupled components may require extensive rework or even complete overhauls when changes are needed. Similarly, unoptimized data handling and redundant components or modules can make maintenance tasks time-consuming and error-prone. 
}

\textit{[MS]}: \faGithub \hspace{0.5mm} In this commit\footnote{\url{https://github.com/pass-culture/pass-culture-main/commit/7c39552d5e76f8d3aadcfebfa14e3a6215bd9ece}}, the developers of the \textit{Pass Culture API} project \textcolor{black}{reused} an architectural solution from SO\footnote{\url{https://stackoverflow.com/a/55900800/12381813}} to improve dependency management and system maintainability. Specifically, they removed the PyMemoize dependency, which was no longer maintained, and re-implemented the caching mechanism using Python's built-in \texttt{lru\_cache()} function to enhance long-term system sustainability: ``\textit{(PC-8124) remove PyMemoize dependency. Its way of importing collections will break with Python 3.9 and the repo is not maintained anymore. We re-implement the caching mechanism with lru\_cache(), inspired by \href{https://stackoverflow.com/a/55900800}{https://stackoverflow.com/a/55900800}}''. 



\textcolor{black}{\textit{\textbf{Reusability issue}} occurs when components or modules of a software system cannot be easily reused in different contexts without significant modification. This typically arises from a lack of modularity, flexibility, or clear abstraction in the system's design, making it difficult to extract and repurpose code for other use cases. Reusability problems often stem from tight coupling, lack of clear interfaces, or components being overly specific to a particular scenario.} 

\textit{[MS]}: \faGithub \hspace{0.5mm} In this commit\footnote{\url{https://github.com/IASI-SAKS/groucho/commit/c530b6aa12c685b0927c70f5c729ee520985fce6}}, the developers of the \textit{GROUCHO} project \textcolor{black}{reused} an architectural solution from SO to configure the \texttt{Maven} plugin to enable the reuse of a module by other modules. This touched on the modular architecture of the system, where different modules need to be independently packaged and reused. Proper configuration of the \texttt{Maven} plugin ensures that the module can be packaged in a way that allows it to be seamlessly imported and used by other modules, which is a core aspect of modular architecture: ``\textit{Added a configuration to spring-boot-maven-plugin. Such configuration is important in order to package the module so that it can be \textit{reused} by other modules. See the answers to: \href{https://stackoverflow.com/questions/61341102/maven-doesnt-find-imported-class-from-another-module-but-intellij-does}{https://stackoverflow.com/questions/61341102/maven-doesnt-find-imported-class-from-another-module-but-intellij-does}}.'' 


\textcolor{black}{\textit{\textbf{Accessibility issue}} arises when a software system is not designed to be accessible to users with disabilities. These issues often stem from factors like lack of support for assistive technologies (e.g., screen readers), poor design choices (e.g., low contrast, non-descriptive labels), or failure to follow accessibility standards (e.g., WCAG). Accessibility problems are typically evident when users with visual, auditory, cognitive, or motor impairments cannot fully interact with the system. Addressing these issues is crucial for ensuring usability for all users and complying with legal requirements.} 

\textcolor{black}{\textit{[S]} \faHandORight \hspace{0.5mm} ``Also handling \textit{accessibility} was another problem (...). While exploring \textit{accessibility defects} in our applications, I just wanted opinions from my colleague developers on dealing with such accessibility hitches, like component (UI design) \textit{accessibility concerns} (...), and SO community users provided some wonderful insights and takeaways regarding this}.'' P47
\color{black}

\begin{table}
\footnotesize
\caption{Categories of architectural problems that are solved by the \textcolor{black}{reused} architectural solutions from SO}
\label{CategoriesOfArchitecturalProblems}
\begin{tabular}{m{6.5cm}m{1.3cm}}
\hline
\textbf{Category}                                       &\textbf{Counts\newline(MS/S)}   \\\hline
Component design issue                                                        
                                                        & 330/173 \\\cline{1-2}
Architectural anti-pattern                                                     
                                                        & 276/105 \\\cline{1-2}                                                        
Security issue                                                     
                                                        & 198/137  \\\cline{1-2} 
                                                        
Component communication issue                                                        
                                                        & 166/102 \\\cline{1-2}                                                       
Performance issue                                                        
                                                        & 59/78\\\cline{1-2}                                                        
Architectural pattern design issue                                                       
                                                        & 39/59 \\\cline{1-2} 
                                                        
Architectural deployment issue                                                         
                                                         & 13/46 \\\cline{1-2}  

Architectural documentation issue                                                        
                                                        & 19/31 \\\cline{1-2} 

Incompatibility issue                                                        
                                                        & 8/35 \\\cline{1-2}                                                        
Reliability issue                                                        
                                                        & 22/7 \\\cline{1-2}
Extensibility issue                                                        
                                                        & 20/2 \\\cline{1-2}

Scalability issue                                                        
                                                         & 11/9 \\\cline{1-2}
Maintainability issue                                                        
                                                         & 7/6 \\\cline{1-2}                                                          
Reusability issue                                                        
                                                          & 6/2 \\\cline{1-2}          
Accessibility issue                                                        
                                                          & 0/4 \\\cline{1-2}
\end{tabular}
\begin{minipage}{9cm} 
\vspace{0.1cm}
\vspace{0.1cm}
\small  \textbf{Notes}: \textit{The column ``Counts (MS/S)'' shows the frequencies of the corresponding category in the mining study (``MS'') and the survey responses (``S'')}.
\end{minipage}
\end{table}

 \noindent\begin{center}
   \begin{tcolorbox}[colback=black!5, colframe=black!20, width=1.0\linewidth, arc=0.5mm, boxrule=0.8pt, left=1mm, right=1mm, top=0.5mm, bottom=0.5mm, boxsep=0.5mm]
              \textcolor{black}{{{
              \textbf{Key Finding of RQ1:} A broad spectrum of architectural problems in OSS development, spanning 15 categories, are solved by \textcolor{black}{reusing} architectural solutions from SO. Notably, the top three categories of architectural problems are \textit{Component design issue}, \textit{Architectural anti-pattern}, and \textit{Security issue}. This insight underscores the critical areas that are problematic for OSS developers, emphasizing the need for tailored solutions in these key domains.
              }}}
    \end{tcolorbox}
 \end{center}

 \subsubsection{Architectural Solutions from SO that are \textcolor{black}{Reused} in OSS Development (RQ2)}\label{ResultsOfRQ2}
Table \ref{CategoriesOfArchitecturalSOlutions} summarizes the architectural solutions from SO that are \textcolor{black}{reused} in OSS development. \textcolor{black}{Note that a single commit or issue may \textcolor{black}{reuse} multiple architectural solutions from SO in OSS development. We counted each architectural solution \textcolor{black}{reused} in a commit or issue separately, and consequently the total frequency in Table \ref{CategoriesOfArchitecturalSOlutions} exceeds the 984 commits and issues analyzed in our study}. Moreover, we observed that some participants mentioned more than one architectural solution that they \textcolor{black}{reused} when solving architectural problems in OSS development. Therefore, the sum of the frequencies from the survey study exceeds the number of the survey participants (i.e., 227 participants) in Table~\ref{CategoriesOfArchitecturalSOlutions}.
 
\textit{\textbf{Architectural refactoring}} is the most frequently \textcolor{black}{reused} type of architectural solutions from SO in OSS development. The solutions in this category provide techniques for restructuring architectures of systems or components in such a way that they do not alter the external behaviors of the systems or components yet improve their internal structures \cite{stal2014refactoring}. Example of such techniques include techniques for decoupling a large codebase into smaller modules or components, techniques for retrofitting an architectural design pattern, techniques for removing cyclic dependencies among modules. We observed that these technical solutions are provided to solve architectural problems, including architectural anti-patterns, maintainability issues, reusability issues, and scalability issues (see Table \ref{CategoriesOfArchitecturalProblems}). 

\textit{[MS]}: \faGithub \hspace{0.5mm} This commit\footnote{\url{https://github.com/hslayers/hslayers-ng/commit/1227b30b4e513ff6ce94b2470d15f28fe9f1f28a}} implements an approach inspired by an architectural solution described in SO post\footnote{\url{https://stackoverflow.com/a/74728818/12381813}}. The solution addresses refactoring the \textit{HSLayers-NG} project structure, optimizing module dependency management, and improving lazy loading behavior to enhance performance, maintainability, and modularity. 

\textit{[S]} \faHandORight \hspace{0.5mm} ``\textit{(...) if it is something similar to program bugs repair, I had a soft spot for interesting refactoring techniques shared on this site (Stack Overflow). I most of the time remove complex and circular dependency problems in our applications by getting rid of unnecessary dependencies between modules or consolidating duplicate modules, where modules are unified into one. These defects make me nervous all the time}.'' P156

\textit{\textbf{Use of frameworks}}. \textcolor{black}{Frameworks provide partial designs with domain-specific services that shape architectural decisions, as described in \cite{SA2012}. They impose architectural constrains, such as communication patterns (e.g., publish-subscribe bus, callbacks), influencing architectural design of a system. Since not all frameworks qualify as architectural solutions (e.g., testing frameworks may not impact overall architecture), our study focuses on standalone frameworks that directly shape a system's architecture. Libraries (e.g., Spring Security) tied to frameworks (e.g., Spring) are excluded from this category to avoid ambiguity. This category of solutions includes frameworks such as Angular, Spring Framework, Django, Flutter, and ASP.NET Core, which address architectural issues like component design, component communication, and security (see Table \ref{CategoriesOfArchitecturalProblems}).} 

\textit{[MS]}: \faGithub \hspace{0.5mm} This issue\footnote{\url{https://github.com/aspnetboilerplate/aspnetboilerplate/issues/6681}} refers to this SO post\footnote{\url{https://stackoverflow.com/q/51027406/12381813}} to explore more effective use of the ASP.NET Core framework in addressing security issues during the development of \textit{ASP.NET Boilerplate} project. 


\textit{[S]} \faHandORight \hspace{0.5mm} ``\textit{I use angular (referring to Stack Overflow)) since it makes the code easy to maintain, reuse, and again it is It is fully extensible and works well with other libraries. Every feature can be modified or replaced to suit your unique development workflow and feature needs. }'' P141

\textit{\textbf{Architectural tactics}}, as design primitives, provide mechanisms for achieving specific quality attribute responses by influencing elements of a quality model through architectural design decisions \cite{harrison2010architecture}. This category encompasses solutions that describe and explain architectural tactic aimed at addressing one specific quality attribute issue in software systems (see Table \ref{CategoriesOfArchitecturalProblems}). Examples include the tactic heartbeat for availability, reduce coupling for maintainability, resource pooling or scheduling for performance, and authentication for security.

\textit{[MS]}: \faGithub \hspace{0.5mm} In this commit\footnote{\url{https://github.com/cloudigrade/cloudigrade/commit/f0c5bbfe6eec40664a79adf768fae6effd8f1f68}}, the developers of the \textit{cloudigrade} project \textcolor{black}{reused} an architectural tactic inspired by discussions from these SO posts: ``\textit{\href{https://stackoverflow.com/questions/55249197/what-are-the-consequences-of-disabling-gossip-mingle-and-heartbeat-for-celery-w}{https://stackoverflow.com/questions/55249197/what-are-the-consequences-of-disabling-gossip-mingle-and-heartbeat-for-celery-w}, \href{https://stackoverflow.com/questions/21132240/celery-missed-heartbeat-on-node-lost}{https://stackoverflow.com/questions/21132240/celery-missed-heartbeat-on-node-lost}, and  \href{https://stackoverflow.com/questions/66978028/application-impacts-of-celery-workers-running-with-the-without-heartbeat-fla}{https://stackoverflow.com/questions/66978028/application-impacts-of-celery-workers-running-with-the-without-heartbeat-fla}}''. The tactic focuses on addressing issues related to the heartbeat mechanism to enhance system reliability and task coordination. 

\textit{\textbf{Use of protocols}}: The solutions in this category gather and clarify the use of various protocols (such as OAuth, HTTPS, TCP, TLS, FTP, SSH, SSL, and SMTP) to address diverse architectural issues, including component communication, security, component design, and reliability issues (see Table \ref{CategoriesOfArchitecturalProblems}). 

\textit{[MS]}: \faGithub \hspace{0.5mm} This commit\footnote{\url{https://github.com/magnusbrok/KabaleSpiller/commit/62d102edb39a35a950a938f1c9b7ab062fef7f50}} refers to this SO post\footnote{\url{https://stackoverflow.com/q/4185242/12381813}} to implement a TCP client in Python for establishing communication with a Java server in the development of the \textit{KabaleSpiller} project. 

 
\textit{\textbf{Use of libraries}}. \textcolor{black}{Libraries typically provide isolated functionalities rather than comprehensive architectural solutions; however, certain libraries, such as React, significantly influence application design and qualify as architectural solutions \cite{albin2003art}. For instance, React's component-based architecture promotes modularity and adheres to principles like separation of concerns and encapsulation, enabling scalable and maintainable UI design. This category encompasses various types of libraries for architecture design and implementation, including those addressing performance challenges (e.g., High Performance Primitive Collections (HPPC), Efficient Java Matrix Library (EJML), Matrix Toolkit Java (MTJ)) and maintainability and reusability concerns (e.g., React). Additionally, we observed that libraries in this category have been \textcolor{black}{reused} to address diverse architectural problems, such as component design issues, incompatibility issues, and component communication issues (see Table \ref{CategoriesOfArchitecturalProblems}).}

\textit{[MS]}: \faGithub \hspace{0.5mm} This commit\footnote{\url{https://github.com/swagger-api/swagger-ui/issues/6996}} refers to this SO post\footnote{\url{https://stackoverflow.com/q/42906358/12381813}} to explore and leverage the \texttt{Redux} library for state management, component design, and application scaling, while maintaining modularity and state isolation in the development of the \textit{Swagger UI} project. The proposed solution demonstrates how the \texttt{Redux} library can be \textcolor{black}{reused} to structure state effectively, ensuring that components are reusable and independent, and preventing interference between their functionalities.

\textit{[S]} \faHandORight \hspace{0.5mm} ``\textit{Well, if I can add my appreciation, I liked this SO site (...). We often use OpenCV in our current projects (...). One of the most interesting projects I've worked on lately is about image processing. The goal was to design a system to be able to recognize Coca-Cola in C++ using the OpenCV library. To name a few, I liked the opinions on Scale-Invariant Feature Transform (SIFT) and Speeded Up Robust Features (SURF) approaches from fellow developers at StackOverflow (...)}.'' P102 

\textit{\textbf{Use of APIs}}. This category of solutions provide and elaborate the usage of APIs (e.g., Java 7's SSLSocket/SSLEngine API, Facebook API, REST API, Graph API, and SOAP API) for architecture design and implementation. OSS developers used the API to address architectural issues, such as component design issues, component communication issues, and incompatibility issues (see Table \ref{CategoriesOfArchitecturalProblems}).

\textit{[MS]}: \faGithub \hspace{0.5mm} In this commit\footnote{\url{https://github.com/nwpushuai/GerritCodeReview/commit/d1e2c2f2c848a042854d38bd0f3c2ec323f47a6b}}, the developers of the \textit{GerritCodeReview} project adopted an architectural solution from this SO post\footnote{\url{https://stackoverflow.com/a/17979954}} to address a security issue by implementing the verify message integrity tactic \cite{SA2012}. The solution involved \textcolor{black}{reusing} Java 7's SSLSocket/SSLEngine API to enhance the system's enforcement of secure communication and mitigate vulnerabilities such as man-in-the-middle attacks.

\textit{[S]} \faHandORight \hspace{0.5mm} ``\textit{(...) Large projects I worked on already had specific development channels for resolving issues; if they did not, I used forums like SO. A little while back, I got help from this community on best practices for structuring the REST API for client-server communications in our Microservice projects (...)}.'' P65

\textit{\textbf{Architectural patterns}} are reusable solutions to common problems in design, defining the structure and behavior of a system \cite{SA2012}. 
This category of solutions provide architectural patterns (e.g., MVC, Broker, Microservice, Client-Server, and Layered pattern) for addressing multiple system quality attributes, such as reusability and scalability. It also explains the usage of architectural patterns for solving architectural problems, such as architectural anti-patterns and component communication (see Table \ref{CategoriesOfArchitecturalProblems}).  


\textit{[MS]}: This issue\footnote{\url{https://github.com/spring-projects/spring-batch/issues/2350}} refers to this SO post\footnote{\url{https://stackoverflow.com/q/3570610/12381813}} for structuring the application in the form of event-driven architectural pattern in the development of the \textit{Spring Batch} project. 


\begin{table}
\footnotesize
\caption{Categories of architectural solutions from SO that are \textcolor{black}{reused} in OSS development}
\label{CategoriesOfArchitecturalSOlutions}
\begin{tabular}{m{6.5cm}m{1.3cm}}
\hline
\textbf{Category}                                         &\textbf{Counts\newline(MS/S)}   \\\hline
Architectural refactoring                                 
                                                          & 554/167 \\\cline{1-2}
                                                          
Use of frameworks                                         
                                                          & 110/133 \\\cline{1-2}
                                                          
Architectural tactic                                      
                                                          & 89/91   \\\cline{1-2}
                                                          
Use of protocols                                          
                                                          & 92/73   \\\cline{1-2}

Use of libraries                                          
                                                          & 79/61   \\\cline{1-2}                                                          

Use of APIs                                               
                                                          & 68/80   \\\cline{1-2}                                                 
                                                                                                                    
Architectural pattern                                     
                                                          & 64/71   \\\cline{1-2} 
\end{tabular}
\end{table}

 \noindent\begin{center}
   \begin{tcolorbox}[colback=black!5, colframe=black!20, width=1.0\linewidth, arc=0.5mm, boxrule=0.8pt, left=1mm, right=1mm, top=0.5mm, bottom=0.5mm, boxsep=0.5mm]
              \textcolor{black}{{{
              \textbf{Key Findings of RQ2}: \textcolor{black}{A variety of architectural solutions (7 categories) from SO are \textcolor{black}{reused} in OSS development. \textit{Architectural refactoring}, \textit{Use of frameworks}, and \textit{Architectural tactic} are the three most frequently employed architectural solutions. This diversity underscores SO's pivotal role as a knowledge-sharing platform for practical architectural solutions.
              }}}} 
    \end{tcolorbox}
 \end{center} 

\color{black}
\subsubsection{Ways of Reusing Architectural Solutions from SO in OSS Development (RQ3)}\label{ResultsOfRQ3}
\color{black}
Practitioners adopted different ways when \textcolor{black}{reusing} architectural solutions from SO in OSS development. In Table~\ref{WaysOfArchitecturalSOlutions}, we present the reported ways in which architectural solutions are \textcolor{black}{reused} in OSS development, organized into five categories. Note that one commit or issue may employ more than one way when \textcolor{black}{reusing} an architectural solution from SO, and consequently, the sum of the frequencies from the mining study exceeds 984 commits and issues in Table \ref{WaysOfArchitecturalSOlutions}. Moreover, since a participant may provide more than one response (way), the sum of the types of way reported in Table \ref{WaysOfArchitecturalSOlutions} exceeds the total number of survey participants. In the following, we describe the reported ways in which architectural solutions are \textcolor{black}{reused} in OSS development.

\textit{\textbf{Convert ideas in the architectural solutions into code}}. Developers transform architectural solutions into code by reusing the ideas in the solutions.

\textit{[MS]}: \faGithub \hspace{0.5mm} In this commit\footnote{\url{https://github.com/opennode/waldur-openstack/commit/1f1cd72787065df4318aaa8701a60fd1a4428313}}, the developers of the \textit{Waldur OpenStack} project implemented an architectural solution based on Django Rest Framework (DRF) and its validation mechanisms, as described in this SO post\footnote{\url{https://stackoverflow.com/q/32834026/12381813}}, incorporating its ideas into the project's development. 

\textit{[S]} \faHandORight \hspace{0.5mm} ``\textit{I am a developer acting as a maintainer, and my job is to code. I just turn the discussed design information (solutions from SO) into code}.'' P82 

\textit{\textbf{Adapt the architectural solutions to fit in the project context}}. Before \textcolor{black}{reusing} architectural solutions from SO, developers make some modifications that may or may not alter the functionality of those architectural solutions in order to fit into the project context. \textcolor{black}{These modifications may involve altering component interfaces or adding supplement components to ensure that the solutions integrate smoothly with the project's existing structure and environment.}

\textit{[MS]}: \faGithub \hspace{0.5mm} In this commit\footnote{\url{https://github.com/thm-projects/arsnova.click/commit/62454ec6ca99fadbeb73519ca1d16c32348ac746}}, the developers of the \textit{arsnova.click} project modified an architectural solution from this SO post\footnote{\url{https://stackoverflow.com/a/45660021}} to make it fit into the project context. The solution defines the interaction between system components, specifically the \texttt{fast-render} package and \texttt{Meteor's Core}. These modifications address low-level incompatibility between the components, resolving problems with their integration and communication at a low level, such as data formats, protocols, or dependencies, to ensure smooth operation within the system.


\textit{[S]} \faHandORight \hspace{0.5mm} ``\textit{Some design solutions (for instance, authentication strategies) are complex and sometimes not provided in technologies similar to mine. I chose to make minor or major modifications when using these solutions, which may improve the simplicity and consequently also make those solutions work in my project technology domains (...)}.'' P58

\textit{\textbf{Convert ideas in the architectural solutions into detail design}}. Before \textcolor{black}{reusing} architectural solutions from SO, developers add more details to the solutions towards components design and their implementations. They define the logical structure of each component and their interfaces to communicate with other components. 

\textit{[MS]}: \faGithub \hspace{0.5mm} In this commit\footnote{\url{https://github.com/uclapi/uclapi/commit/d4c70201cca42270ab85d684356be2c453fbc563}}, developers transformed the architectural solutions from this SO post\footnote{\url{https://stackoverflow.com/a/11024387/12381813}} and this SO post\footnote{\url{https://stackoverflow.com/a/62450640/12381813}} into detained designs. Specifically, the developers provided detailed designs for mitigating Cross-Site Request Forgery (CSRF) attacks in Web application architectures. This included illustrating XHR-based CSRF contexts, design considerations, mitigation strategies, and the separation of concerns between front-end and back-end. Furthermore, they clarified OAuth authentication, JWT session management, SameSite cookies, and CSRF protection while aligning application logic refactoring with the goals of the \textit{UCL API} project.



\textit{\textbf{Learn the architectural knowledge in the architectural solutions}}. Developers indirectly \textcolor{black}{reuse} architectural solutions from SO. Specifically, they treat the SO posts that provide architectural solutions as a knowledge source related to the architectural issues they are addressing. 

\textit{[MS]}: \faGithub \hspace{0.5mm} In this issue\footnote{\url{https://github.com/supabase/auth/issues/114}}, developers did not directly \textcolor{black}{reuse} an architectural solution from SO. But instead they referenced an architectural solution from this SO post\footnote{\url{https://softwareengineering.stackexchange.com/a/408592/422568}} discussing an architectural issue that is contextually similar to the issue reported in the development of the \textit{Supabase} project. 


\textit{[S]} \faHandORight \hspace{0.5mm} ``\textit{While explaining the design issue and its potential solution, and enrich my knowledge about certain concepts as well, I can also add a link to the post to show my respect (appreciation) to the original post author}.'' P33.

\textit{\textbf{\textcolor{black}{Reuse} the architectural solutions directly in the project}}. Developers copy and paste architectural solutions from SO without any modification. 

\textit{[S]} \faHandORight \hspace{0.5mm} ``\textit{Rarely, I directly use the solution as it is provided, but, I say not often though :)}.'' P203

\begin{table}
\footnotesize
\caption{Ways of architectural solutions utilization from SO in OSS development}
\label{WaysOfArchitecturalSOlutions}
\begin{tabular}{m{7cm}m{1cm}<{\centering}}
\hline
\textbf{Category}                                       &\textbf{Counts\newline(MS/S)}    \\ \hline
Convert ideas in the architectural solutions into code    
                                                        & 524/211 \\\cline{1-2}
Adapt the architectural solutions to fit in the project context
                                                        & 227/133 \\\cline{1-2} 
Convert ideas in the architectural solutions into detail design
                                                        & 145/57 \\\cline{1-2}                                                         
Learn the architectural knowledge in the architectural solutions
                                                        & 85/42 \\\cline{1-2}                                                       
\textcolor{black}{Reuse} the architectural solutions directly in the project  
                                                        & 16/41 \\\cline{1-2}                   
\end{tabular}
\end{table}


\noindent\begin{center}
   \begin{tcolorbox}[colback=black!5, colframe=black!20, width=1.0\linewidth, arc=0.5mm, boxrule=0.8pt, left=1mm, right=1mm, top=0.5mm, bottom=0.5mm, boxsep=0.5mm]
              \textcolor{black}{{{
              \textbf{Key Findings of RQ3}: OSS developers often rely on ad hoc ways (e.g., informal, improvised, or unstructured approaches) to \textcolor{black}{reuse} architectural solutions from SO, drawing on personal experience and intuition rather than standardized or systematic practices. Key methods observed include \textit{converting ideas in the architectural solution into code} and \textit{adapting the architectural solutions to fit the project context}, highlighting that OSS developers view SO as a reference of architectural solutions that can be flexibly applied to meet the specific needs of a project. 
              }}} 
    \end{tcolorbox}
 \end{center}

\subsection{RQs Answered by the Survey Study}\label{ResultsOfRQ4_to_RQ5}
As stated in Section \ref{Methodology}, the last two RQs (i.e., RQ4 and RQ5) are particularly answered by the data gathered from the survey study (see Section \ref{CollectingSurveyData}). Thus, this section reports the results for RQ4 and RQ5 obtained from 227 valid survey responses. 

\color{black}
\subsubsection{Characteristics of Reused Architectural Solutions and Their Descriptions in OSS Development (RQ4)}\label{ResultsOfRQ4}
\color{black}
To investigate RQ4, we asked survey participants to describe the characteristics of architectural solutions from Stack Overflow (SO) that they consider when reusing those solutions in OSS development (i.e., SQ9). The reported characteristics are summarized by frequency in Table~\ref{CharacteristicsOfArchitecturalSOlutionsFromSOInOSSDevlpmt}. \textcolor{black}{We further organized these characteristics into two main categories: (1) \textit{Characteristics of architectural solutions}, which refer to the inherent attributes or outcomes of applying the solutions, regardless of how they are described; and (2) \textit{Characteristics of solution descriptions}, which concern how the architectural solution is presented in the post, including aspects such as clarity, level of technical details, and inclusion of code samples and diagrams.} Note that participants could mention multiple characteristics, so the total number of reported characteristics in Table~\ref{CharacteristicsOfArchitecturalSOlutionsFromSOInOSSDevlpmt} exceeds the number of survey respondents.

\color{black}
\begin{table}[h]
\footnotesize
\caption{\textcolor{black}{Characteristics of reused architectural solutions and their descriptions from SO in OSS development}}
\label{CharacteristicsOfArchitecturalSOlutionsFromSOInOSSDevlpmt}
\begin{tabular}{p{7cm}r}
\hline
\textbf{\textcolor{black}{Characteristics of Solution Descriptions}} & \textbf{Count} \\
\hline
The architectural solution description includes a sample of code & 177 \\
The architectural solution description provides the design context & 154 \\
The architectural solution description includes a concise summarization & 97 \\
The architectural solution description provides the source of the solution & 64 \\
The architectural solution description includes an architectural diagram (e.g., component diagram) & 38 \\
The architectural solution is accepted by the asker of the question & 17 \\
\hline
\textbf{\textcolor{black}{Characteristics of Architectural Solutions}} & \textbf{} \\
\hline
The architectural solution makes the system easier to extend and maintain & 5 \\
The architectural solution positively impacts certain quality attributes of the system & 3 \\
Amount of time available to utilize an architectural solution during development & 2 \\
\hline
\end{tabular}
\end{table}
\color{black}

\textbf{\textcolor{black}{a) Characteristics of Solution Descriptions}}

\textbf{The architectural solution description includes a sample of code}. OSS developers noted that they prefer SO architectural solutions that are accompanied by a code sample, as it demonstrates how the solutions can be practically implemented.  

\textit{[S]} \faHandORight \hspace{0.5mm} ``\textit{As a developer, my first job is to code the confirmed solutions, thus, with no doubt the design solution with few and important code (main components/classes or modules, message transmission protocols) is of the importance for me.'' P108}

\textbf{The architectural solution description provides the design context}. OSS developers also emphasized their preference for SO architectural solutions that include context about the design, such as the problems being addressed and the rationale behind the solutions.

\textcolor{black}{\textit{[S]} \faHandORight \hspace{0.5mm} ``\textit{I check how well the solution aligns  within my system's components/classes or context:-)}.'' P93}

\textbf{The architectural solution description includes a concise summarization}. Some OSS developers  expressed that they favor architectural solutions that offer a clear and concise summary, providing a quick overview of the key aspects of the solutions. 

\textcolor{black}{\textit{[S]} \faHandORight \hspace{0.5mm} ``\textit{Lengthy solutions require too much time for me to grasp the main ideas, I go for succinct solutions first for efficiency and clarity}.'' P219}

\textbf{The architectural solution description provides the \textcolor{black}{source} of the solution}. Some OSS developers mentioned that they like architectural solutions that explain the origin or source, including the development history and any relevant references.

\textcolor{black}{\textit{[S]} \faHandORight \hspace{0.5mm} ``\textit{Intermittently, I ask for the source of the answer before I use it for the sake of ensuring its credibility and reliability}.'' P157}

\textbf{The architectural solution description includes an architectural diagram (e.g., component diagram)}. Some OSS developers noted that they find architectural solutions with accompanying architectural diagrams helpful, as these diagrams visually represent the architecture and clarify the relationships between components.

\textcolor{black}{\textit{[S]} \faHandORight \hspace{0.5mm} ``\textit{I may value answers that include diagrams, as they help clarify concepts that might otherwise be ambiguous and require making multiple assumptions to understand}.'' P117}

\textbf{The architectural solution is accepted by the asker of the question}. Some OSS developers indicated that architectural solutions accepted by the asker are seen as more reliable or trustworthy, as these solutions often reflect the solution that best addresses the original problem.

\textcolor{black}{\textit{[S]} \faHandORight \hspace{0.5mm} ``\textit{If it is accepted and liked by the community, then it gives me some confidence in its applicability to my project}.'' P163}

\textbf{\textcolor{black}{b) Characteristics of Architectural Solutions}}

\textbf{The architectural solution makes the system easier to extend and maintain}. Some OSS developers emphasized that architectural solutions that facilitate system long-term maintainability are highly valued, as they contribute to smoother future development and easier updates.

\textcolor{black}{\textit{[S]} \faHandORight \hspace{0.5mm} ``\textit{I won't opt for complex solutions, but rather flexible ones that can support future requirements (...) while minimizing maintenance overhead, as they are more likely to align with long-term goals of my project}.'' P54}


\textbf{The architectural solution positively impacts certain quality attributes of the system}. Few OSS developers noted that architectural solutions which enhance key quality attributes, such as performance, security, or reliability, are particularly valued for the long-term benefits of the system.

\textcolor{black}{\textit{[S]} \faHandORight \hspace{0.5mm} ``\textit{Who wouldn't want to improve their project? The goal of our development team is always to enhance system quality, so solutions should contribute to and elevate the overall quality of the system, etc}.'' P194}

\textbf{Amount of time available to utilize an architectural solution during development}. Finally, few OSS developers mentioned that the time available to implement an architectural solution is a critical factor, as limited time may lead to trade-offs in the depth or quality of the solution applied.

\textcolor{black}{\textit{[S]} \faHandORight \hspace{0.5mm} ``\textit{I also consider the amount of resources (e.g., time, etc) available to incorporate the answer (specifically, if the solution is more complex) in my software project}.'' P91}
\color{black}

 
 \noindent\begin{center}
   \begin{tcolorbox}[colback=black!5, colframe=black!20, width=1.0\linewidth, arc=0.5mm, boxrule=0.8pt, left=1mm, right=1mm, top=0.5mm, bottom=0.5mm, boxsep=0.5mm]
              \textcolor{black}{{{
              \textbf{Key Findings of RQ4}:  Practitioners consider a range of characteristics (9 types) of architectural solutions from SO when reusing them in OSS development. \textcolor{black}{These characteristics are grouped into two main categories: Characteristics of Solution Descriptions and Characteristics of Architectural Solutions}. Among these, \textit{the architectural solution description includes a sample of code} and \textit{the architectural solution description provides the design context} are the most frequently mentioned. These findings highlight the importance of both practical and contextual information in adopting external architectural solutions. The frequent mention of code samples indicates that OSS developers prioritize solutions with clear, implementable examples, which streamline their integration into ongoing projects. 
              }}}          
    \end{tcolorbox}
\end{center}

\subsubsection{Challenges Faced when \textcolor{black}{Reusing} Architectural Solutions from SO in OSS Development (RQ5)}\label{ResultsOfRQ5}\color{black}

To examine RQ5, we asked our participants to write about their challenges when \textcolor{black}{reusing} architectural solutions from SO in OSS development (i.e., SQ11). The main challenges are listed according to their counts in Table \ref{ChallengesFacedWhenUtilizingArchitecturalSOlutions}. Note that since a participant may provide more than one response (challenge), the sum of the types of the challenges reported in Table \ref{ChallengesFacedWhenUtilizingArchitecturalSOlutions} exceeds the total number of survey participants. 

\begin{table}
\footnotesize
\caption{Challenges faced when \textcolor{black}{reusing} architectural solution from SO in OSS development}
\label{ChallengesFacedWhenUtilizingArchitecturalSOlutions}
\begin{tabular}{m{7.2cm}m{0.7cm}<{\centering}}
\hline
\textbf{Category}                                      &\textbf{Count}   \\ \hline
\textcolor{black}{Significant time is required to adapt the architectural solution from SO to address design concerns raised in OSS development}  & \textcolor{black}{170}  \\\cline{1-2}
The architectural solution from SO is not provided in the design context of the given OSS project    & 115  \\\cline{1-2} 
Figuring out the up-to-dateness (e.g., latest technologies) of the architectural solution            & 96   \\\cline{1-2}
Judging the credibility of the architectural solution                                                & 59   \\\cline{1-2}
Contradictions (e.g., both positive and negative impacts on certain quality attributes) in the architectural solution description  & 39 \\\cline{1-2} 
Finding the architectural solution that addresses design concerns (e.g., scalability) raised in OSS development  & 6 \\\cline{1-2}
\textcolor{black}{Time-consuming implementation of architectural solutions from SO in OSS development}  & \textcolor{black}{4} \\\cline{1-2}
Readability issues in the architectural solution description                                          & 3    \\\cline{1-2}
\end{tabular}
\end{table}

\textcolor{black}{\textbf{Significant time is required to adapt the architectural solution from SO to address design concerns raised in OSS development}}. This challenge arises because architectural solutions on SO are often tailored to specific scenarios. Developers must adapt these solutions to their OSS projects' unique requirements and constraints, often facing integration issues with existing codebases that use different architectural styles, design patterns, or technology stacks. This process can lead to unforeseen challenges, demanding considerable time and effort for adjustments.

\textcolor{black}{\textit{[S]} \faHandORight \hspace{0.5mm} ``\textit{A solution on Stack Overflow may work in isolation, but adapting it to address specific design concerns raised in the project, while aligning it with our design patterns, components, and overall structure, often requires significant time.}'' P114}

\textcolor{black}{\textbf{The architectural solution from SO is not provided in the design context of the given OSS project}. Another recurring challenge identified from the survey responses is the mismatch or lack of design context in architectural solutions on SO. For example, SO posts often omit details about the specific project or problem they address, sometimes due to privacy concerns, making it difficult for developers to align the solutions with the specific architecture or design goals of their OSS projects, leading to confusion and potential misapplication.} 

\textcolor{black}{\textit{[S]} \faHandORight \hspace{0.5mm} ``\textit{Many thanks for Stack Overflow, it sometimes saves my time. I can find several workarounds for my design issues there, however, the context directs the use of such workarounds/solutions, we mainly face this issue}.'' P55} 

\textcolor{black}{\textbf{Figuring out the up-to-dateness (e.g., latest technologies) of the architectural solution}. The fast-evolving tech landscape introduces challenges for OSS developers, as architectural solutions on SO may become outdated with new design patterns, frameworks, APIs, and best practices. Furthermore, while SO is a community-driven platform, not all the solutions are actively maintained or reviewed, making it difficult for developers to gauge their relevance or alignment with recent technological advancements.}

\textcolor{black}{\textit{[S]} \faHandORight \hspace{0.5mm} ``\textit{Some posts (on SO) are several years old and may suggest archaic technologies or approaches, potentially resulting in quality problems like performance issues (...)}.'' P203}

\textcolor{black}{\textbf{Judging the credibility of the architectural solution}. Architectural solutions on SO vary in credibility due to contributors' differing expertise levels. OSS developers often struggle to distinguish seasoned professionals' insights from less experienced users. The absence of supporting references or authoritative links further complicates validating solutions, leading to uncertainty about their reliability.}

\textcolor{black}{\textit{[S]} \faHandORight \hspace{0.5mm} ``\textit{I occasionally have concerns about the credibility of the answers (on Stack Overflow) to my queries when using those answers in my project to achieve the desired outcomes}.'' P120}

\textcolor{black}{\textbf{Contradictions (e.g., both positive and negative impacts on certain quality attributes) in the architectural solution description} on SO create challenges for OSS developers in evaluating and \textcolor{black}{reusing} these solutions. Variability in experiences, contextual differences, and subjective assessments of quality attributes, such as performance and security, contribute to this complexity. Developers’ differing priorities, such as favoring performance over security or vice versa, can lead to conflicting evaluations of the same solution, necessitating careful analysis and utilization.}

\textcolor{black}{\textit{[S]} \faHandORight \hspace{0.5mm} ``\textit{Some answers on Stack Overflow may contain conflicting and contradictory discussions on the usefulness of the answers, which can lead to confusion for me when determining the most appropriate solution for my specific needs}.'' P131}

\textcolor{black}{\textbf{Finding the architectural solution that addresses design concerns (e.g., scalability) raised in OSS development}. With a vast amount of content available on SO, OSS developers struggle to sift through numerous architectural solutions and discussions to find those that specifically address their design concerns. Moreover, searching for these solution on SO can be challenging, as the search algorithms may not effectively prioritize solutions relevant to specific design concerns \cite{nadi2020essential}. This can result in developers missing out on valuable insights that could address their needs.}

\textcolor{black}{\textit{[S]} \faHandORight \hspace{0.5mm} ``\textit{It is difficult for me to find suitable design solution (like caching techniques) based on my particular needs, such as performance, reliability, language, etc}.'' P111}

\textcolor{black}{\textbf{Time-consuming implementation of architectural solutions from SO in OSS development}.
This category captures challenges related to the considerable time and effort required to implement architectural solutions from SO into OSS projects. Even when a solution appears suitable at a conceptual level, turning it into working code for a specific project context can be demanding and time-intensive.}

\textcolor{black}{\textit{[S]} \faHandORight \hspace{0.5mm} ``\textit{One big problem is the time it takes to transform what is written in the design solution into the code I want for my project domain.}''P1}

\textcolor{black}{\textbf{Readability issues in the architectural solution description}. 
Architectural solutions on SO may contain complex terminologies, concepts, and jargons that are not clearly written and explained, making it difficult for OSS developers to read, understand, and apply them effectively. 
Additionally, issues like spelling, grammar, and punctuation errors in architectural solutions description on SO can detract from the professionalism of the response. Such errors also impede readability and comprehension, distracting readers from the core message and reducing the effectiveness of the solution.}

\textcolor{black}{\textit{[S]} \faHandORight \hspace{0.5mm} ``\textit{Readability is an issue in some answers on SO:)}.'' P73} 

\noindent\begin{center}
   \begin{tcolorbox}[colback=black!5, colframe=black!20, width=1.0\linewidth, arc=0.5mm, boxrule=0.8pt, left=1mm, right=1mm, top=0.5mm, bottom=0.5mm, boxsep=0.5mm]
              \textcolor{black}{{{
             \textbf{Key Findings of RQ5}: Participants acknowledged two major challenges in \textcolor{black}{reusing} architectural solutions from SO for OSS development: \textit{Significant time is required to adapt the architectural solution from SO to address design concerns raised in OSS development} and \textit{the architectural solution from SO is not provided in the design context of the given OSS project}. These findings emphasize the difficulty of tailoring external architectural solutions to fit OSS requirements and the absence of design-specific context, which complicates effective integration of these solutions. 
        }}} 
    \end{tcolorbox}
 \end{center}

\color{black}
\section{Implications}\label{Implications}
\textcolor{black}{This section first discusses the mappings between architectural solutions and problems, and then the implications of our study results for various stakeholder groups.}

\color{black}
\subsection{Architectural Solutions to Problems Mapping}\label{Mapping_Problems_And_Solutions}

Table~\ref{tab_problem_solution_mapping} presents the mapping between architectural problem categories encountered in OSS projects and the architectural solution categories reused by developers. Abbreviations (e.g., “AR” for \textit{Architectural Refactoring}) are used for each solution category, with full names provided in the table’s footnote. The matrix reveals noteworthy trends:
 
\textit{Dominance of Architectural Refactoring (AR)}: Across nearly all problem categories, AR emerges as the most frequently adopted solutions. This is particularly evident in resolving: Architectural Anti-patterns (198 instances), Security Issues (101), Component Design Issues (208), and to a lesser extent, Component Communication Issues (11). This suggests that developers often prefer to restructure the existing architecture in order to resolve deep-rooted issues without altering functionality.

\textit{Targeted Use of Frameworks (UF) and APIs (UA}): Use of Frameworks (UF) is prominently applied to address Component Communication Issues (45) and Security Issues (24), indicating the role of established frameworks in facilitating communication protocols and enforcing security standards. UA is also leveraged significantly in Component Communication (45) and modestly across other categories, reflecting the modularization benefits and integration facilitation provided by APIs.

\textit{Tactics (AT) and Protocols (UP) as Niche Solutions}: Architectural Tactics (AT) are primarily applied to address Performance Issues (40), Security Issues (26), and Reliability Issues (7), consistent with the literature where tactics are employed to achieve specific quality attributes~\cite{SA2012}. Use of Protocols (UP) is more diffusely applied, with moderate usage across Security, Component Communication, and Extensibility concerns, indicating developers' preference for standardized protocols that facilitate secure and maintainable architectures, particularly in distributed systems.

\textit{Negligible Solutions for Some Issues}: Some problem categories, such as Architectural Documentation Issues and Incompatibility issues, are weakly addressed, highlighting areas where SO falls short in bridging the gap between architectural problems and their corresponding solutions.

\textit{Solution Diversity by Problem Categories}: Problem categories such as Component Design and Security issues are associated with a diverse set of solutions, suggesting complex, multifaceted challenges requiring multi-pronged strategies. In contrast, Architectural Anti-patterns are mainly addressed through refactoring and pattern applications, indicating more straightforward corrective paths.

\begin{table}[ht]
\scriptsize
\setlength{\tabcolsep}{4pt} 
\centering
\caption{\textcolor{black}{Mapping between architectural problem categories (vertical) and solution categories (horizontal)}.}
\begin{tabular}{lccccccc}
\toprule
\textbf{Problem Category} & \textbf{AR} & \textbf{UF} & \textbf{AT} & \textbf{UP} & \textbf{UL} & \textbf{UA} & \textbf{AP} \\
\midrule
Component design issue & \cellcolor{blue!25}208 & \cellcolor{blue!15}10 & 0 & \cellcolor{blue!10}13 & \cellcolor{blue!30}65 & \cellcolor{blue!10}15 & 0 \\
Architectural anti-pattern & \cellcolor{blue!40}198 & 0 & 0 & 0 & 0 & 0 & \cellcolor{blue!20}24 \\
Security issue & \cellcolor{blue!30}101 & \cellcolor{blue!10}24 & \cellcolor{blue!20}26 & \cellcolor{blue!15}24 & 0 & 0 & \cellcolor{blue!10}4 \\ 
Component communication issue & \cellcolor{blue!20}11 & \cellcolor{blue!25}45 & 0 & \cellcolor{blue!15}26 & \cellcolor{blue!10}5 & \cellcolor{blue!25}39 & \cellcolor{blue!20}14 \\
Performance issue & \cellcolor{blue!10}1 & \cellcolor{blue!5}12 & \cellcolor{blue!25}40 & \cellcolor{blue!5}2 & 0 & 0 & \cellcolor{blue!5}4 \\
Architectural pattern design issue & \cellcolor{blue!15}4 & 0 & 0 & \cellcolor{blue!5}0 & \cellcolor{blue!10}1 & \cellcolor{blue!5}0 & 0 \\
Architectural deployment issue & \cellcolor{blue!10}3 & 0 & 0 & \cellcolor{blue!5}0 & 0 & \cellcolor{blue!5}1 & 0 \\
Architectural documentation issue & 0 & 0 & 0 & \cellcolor{blue!10}2 & \cellcolor{blue!10}1 & \cellcolor{blue!15}1 & \cellcolor{blue!10}1 \\
Incompatibility issue & \cellcolor{blue!5}2 & \cellcolor{blue!5}1 & 0 & \cellcolor{blue!5}1 & \cellcolor{blue!5}0 & \cellcolor{blue!5}0 & 0 \\
Reliability issue & \cellcolor{blue!10}4 & 0 & \cellcolor{blue!10}7 & \cellcolor{blue!10}2 & 0 & 0 & \cellcolor{blue!5}1 \\
Extensibility issue & \cellcolor{blue!10}6 & \cellcolor{blue!10}4 & \cellcolor{blue!5}2 & \cellcolor{blue!10}4 & 0 & 0 & \cellcolor{blue!5}1 \\
Scalability issue & \cellcolor{blue!5}3 & 0 & \cellcolor{blue!10}3 & 0 & 0 & 0 & \cellcolor{blue!5}1 \\
Maintainability issue & \cellcolor{blue!5}2 & \cellcolor{blue!15}5 & \cellcolor{blue!5}1 & 0 & 0 & 0 & \cellcolor{blue!10}2 \\
Reusability issue & \cellcolor{blue!5}2 & 0 & \cellcolor{blue!5}2 & 0 & 0 & 0 & \cellcolor{blue!5}1 \\
\bottomrule
\end{tabular}
\vspace{2mm}

\footnotesize{
\textbf{Full names of architectural solution categories:}
\textit{\textbf{AR}: Architectural Refactoring}; 
\textit{\textbf{UF}: Use of Frameworks}; 
\textit{\textbf{AT}: Architectural Tactic}; 
\textit{\textbf{UP}: Use of Protocols}; 
\textit{\textbf{UL}: Use of Libraries}; 
\textit{\textbf{UA}: Use of APIs}; 
\textit{\textbf{AP}: Architectural Pattern}.
}
\label{tab_problem_solution_mapping}
\end{table}
\color{black}




\subsection{Implications for Researchers}\label{Implications_for_Researchers}
\textbf{Investigating the potential relationship and mismatch between the architectural problems and solutions}.
The results of RQ1 and RQ2 reveal the broad spectrum of architectural problems in OSS development and the diverse architectural solutions employed to address them. This highlights SO's critical role as a knowledge-sharing platform for practical and adaptable solutions. The alignment between common issues (e.g., \textit{component design issues} and \textit{architectural anti-patterns}) and their corresponding solutions (e.g., \textit{architectural refactoring} and \textit{use of frameworks}) demonstrates SO's relevance in tackling real-world architectural design issues. However, there may be instances of potential mismatches, where the proposed architectural solutions are either unsuitable or incorrectly applied to specific architectural problems. For example, in the case of a scalable architecture issue, where there is a need to handle a large volume of incoming user requests by efficiently distributing them across servers to maintain performance under high load, a solution from SO might propose using a single monolithic server to handle all requests. This would be a mismatch because a single server would likely become a bottleneck, failing to address the scalability issue effectively. Additionally, some solutions may be too narrow or incomplete, addressing only one aspect of a problem while neglecting other critical concerns. For instance, while scalability issues in distributed systems are often addressed with solutions like load balancing techniques, many solutions shared on SO overlook other important factors, such as fault tolerance or latency optimization, key aspects of scalability in certain architectural contexts. 
Future research could explore these mismatches and incomplete solutions to help practitioners better navigate architectural problem-solution pairs and identify areas where SO may fall short in bridging the gap between architectural issues and solutions. A practical approach could involve developing an automatic mapping tool that categorizes architectural problems and their associated solutions shared on SO, offering developers clearer guidance in identifying solutions tailored to their specific architecture design issues.

\textbf{Approaches for converting architectural solutions from SO into code}. 
The results of RQ3 highlight that OSS developers most commonly convert architectural solutions from SO into code. A likely reason is that many OSS practitioners are developers focused on implementation, converting architectural decisions (e.g., selected solutions) directly into their projects. As one survey participant stated, “\textit{I am a developer acting as a maintainer, and my job is to code. I just turn the discussed design information (solutions from SO) into code}” (P82). However, manually converting architectural solutions into code is time-consuming and labor-intensive. Recent advancements in Large Language Models (LLMs) have shown promise in automating various software engineering tasks \cite{zheng2023towards}. Therefore, future research could explore LLM-based approaches (e.g., retrieval-augmented generation \cite{fan2024survey} with external knowledge of architecture to code) that assist developers in automating the conversion of architectural solutions from SO into code or generating sample code, ultimately streamlining the architecture implementation in OSS projects.

\textbf{Evaluating the quality of architectural solutions from SO}. While OSS developers frequently leverage architectural solutions from SO, the quality of these solutions remains uncertain. As one participant noted, “\textit{Sometimes the quality of the solution is not enough. I need to adapt the solution and improve its performance}” (P192). Previous studies have identified several risks associated with solutions from SO, including code reuse violations \cite{an2017stack}, security vulnerabilities \cite{fischer2017stack}, and obsolescence \cite{obsolete2019}. Similarly, adopting unverified architectural solutions without thorough evaluation can introduce vulnerabilities, inefficiencies, and system instability. By assessing these solutions, developers can make more informed decisions, minimizing rework and enhancing the overall quality of OSS systems. Additionally, identifying potential pitfalls, such as missing design context or platform-specific considerations, can highlight areas for improvement in SO's offerings, ultimately leading to more stable and scalable architectures. Given the critical role of architecture in OSS projects \cite{ding2014open, bi2021architecture}, it is essential to rigorously evaluate the quality, relevance, and applicability of architectural solutions from SO. 

\textbf{Assessing the up-to-dateness and credibility of architectural solutions on SO}. The results of RQ5 also reveal that the survey participants encountered difficulties in determining the up-to-dateness (e.g., outdated software protocols) and credibility of architectural solutions on SO (see Table \ref{ChallengesFacedWhenUtilizingArchitecturalSOlutions}). To address these challenges, researchers could develop an automated evaluation framework that leverages sources like SO and official technical documentation to systematically assess the credibility and timeliness of architectural solutions shared on Q\&A sites. Additionally, an integrated validator could improve the quality of software artifacts by providing an efficient way to test and validate elements like source code. Several Q\&A sites already offer such features. For instance, SO’s validation functionality supports Web languages such as HTML, CSS, and JavaScript. Building on this concept, researchers could propose an architectural solution validator, which can detect outdated architectural solutions and alert users about deprecated design patterns, frameworks, and protocols. Such a validator would ensure that solutions adopted from SO remain current and credible, ultimately supporting OSS developers in creating more robust and maintainable systems. 

\textbf{Further investigation on adaptation of architectural solutions from SO in OSS development}. The results of RQ3 reveal that developers adopt various approaches to integrate architectural solutions from SO, with adapting these solutions to fit the project context being the second most common approach. However, the practical aspects of how developers adapt architectural solutions from SO remain unclear. For instance, it is not well understood whether developers assess multiple facets of these solutions, such as functionality, quality attributes, and dependencies, alongside project-specific factors like codebase structure (e.g., intra-file organization) and the overarching architecture of the OSS project. Patterns of adaptation, including component wiring, component configuration, component invocation, data caching, and API/interface interactions, have also not been thoroughly explored. Understanding these patterns is crucial for uncovering how developers manage integration and customization during adaptation. This includes their workflows and decision-making processes when balancing trade-offs between solution characteristics and project constraints. Investigating these aspects would provide empirical evidence on developers' habits and workflows for adapting architectural solutions from Q\&A sites in OSS development, highlighting current challenges and areas for improvement. Future research could also explore how these adaptation practices impact long-term quality attributes such as maintainability and extensibility. By shedding light on these complexities, researchers could help developers adopt external architectural solutions more effectively, thereby fostering better architecting practices in OSS projects.

\subsection{Implications for Architects and Developers}

\textbf{Referencing the architectural solutions after reuse}. Many developers neglect to reference the sources of architectural solutions they incorporate into OSS development. For instance, during our data filtering, we observed cases where OSS developers mentioned consulting SO to address certain issues but failed to explicitly reference the relevant SO posts. Similarly, Baltes \textit{et al}. \cite{baltes2019usage} highlighted the licensing challenges that arise when reusing SO snippets in OSS projects. Without proper referencing, the history of reuse and adaptation of architectural solutions from SO becomes difficult to trace, potentially resulting in issues such as the proliferation of vulnerabilities or non-compliance with licensing terms. Thus, we recommend that OSS developers reference architectural solutions from SO to: (i) Ensure Traceability: Facilitate tracking the origin and evolution of reused solutions, enabling better maintenance and updates. (ii) Mitigate Risks: Reduce the risk of legal and security issues, such as licensing violations or inadvertently introducing vulnerabilities. 
(iii) Support Reusability: Provide future developers with insights into the rationale and implementation of the adopted architectural solutions, streamlining adaptation in similar contexts.

\subsection{Implications for SO Users and Owners}
\subsubsection{For SO Users}
\textbf{Guidelines for posting architectural solutions}. As revealed by the results of RQ4, OSS developers consider various distinct characteristics of architectural solutions from SO when selecting and \textcolor{black}{reusing} these solutions in their projects. The diversity of these characteristics underscores the differing needs of developers when it comes to the descriptions of architectural solutions. Moreover, the results of RQ4 highlight the significance of both practical and contextual information in adopting external architectural solutions. The frequent mention of code samples in architectural solutions highlights OSS developers' preference for clear and implementable examples that simplify integration process of these solutions into projects. Emphasizing design context further underscores the need to understand the architectural principles behind solutions. Together, these insights suggest that effective architectural solutions from SO should combine actionable code with relevant context to ensure proper application of the solutions in OSS projects. To support developers, SO users can refer to the characteristics in Table \ref{CharacteristicsOfArchitecturalSOlutionsFromSOInOSSDevlpmt} as a guide when sharing architectural solutions. Below, we provide guidelines to help SO users create more applicable and valuable solutions for OSS development, increasing their adoption and utility. 

\textit{Write concise architecture solutions with sample of code and/or diagrams}: SO users are encouraged to provide concise architectural solutions by focusing on key points and avoiding unnecessary details. Including architecturally relevant code samples, when appropriate, can help clarify the solution's implementation. Additionally, users are advised to include architectural diagrams, if needed, to illustrate aspects such as the implementation view or overall structure, making the solution easier to understand and apply. 

\textit{Include \textcolor{black}{source} and contextual details in architectural solution descriptions}: SO users should consider including the \textcolor{black}{sources} of their architectural solutions in the solution description, such as relevant URLs or references, to help OSS practitioners evaluate the credibility and origins of the information. Furthermore, providing design contexts is strongly recommended, as these are crucial for practitioners to adapt and apply the architectural solutions effectively in their specific OSS development projects. 

\subsubsection{For SO Owners}
\textbf{Improving the organization of architectural solutions on SO}. Navigating SO effectively has been reported as a challenge, with concerns about how well it directs users to relevant development information, such as solutions \cite{nadi2020essential}. To improve this, SO owners could enhance the organization and description of architectural solutions by leveraging the characteristics identified in Table \ref{CharacteristicsOfArchitecturalSOlutionsFromSOInOSSDevlpmt}. These improvements would help OSS developers more easily locate and apply solutions suited to their specific needs. For example, instead of solely ranking answers based on their score and creation time \cite{zhang2021comments, nadi2020essential}, SO could prioritize solutions more effectively. This could include implementing filtering and ranking mechanisms that emphasize solutions with explicit design context or those addressing key system qualities. Additionally, enabling users to vote on specific aspects of the solutions, such as relevance to security, performance, or maintainability, could further enhance the quality and utility of architectural solutions. \textcolor{black}{Building on the mismatch issues discussed in Section \ref{Implications_for_Researchers}, SO owners/moderators could explore guidelines for users when evaluating architectural solutions that may appear off-topic or incorrect at first glance. For instance, before downvoting, moderators could apply a simple validation step to assess whether an architectural solution, while not directly aligned with the original question, still offers technically sound guidance. Such guidelines could help preserve potentially valuable architectural knowledge that might otherwise be overlooked.}
 

\subsection{Implications for Tool Designers}
\textbf{Innovative tools for searching and reusing architectural solutions from SO in OSS development}. The results of RQ2 (see Table \ref{CategoriesOfArchitecturalSOlutions} in Section \ref{ResultsOfRQ2}) reveal that architectural refactoring is the most frequently \textcolor{black}{reused} solution from SO for addressing architectural problems in OSS development. This underscores the critical need for OSS developers to restructure system architectures for enhancing maintainability without altering their external behavior \cite{stal2014refactoring}. Furthermore, the seven identified categories of architectural solutions from SO (see Table \ref{CategoriesOfArchitecturalSOlutions}) reaffirm the conclusions of Soliman \textit{et al}. \cite{soliman2017developing} and our previous work \cite{de2023characterizing} that SO serves as a valuable repository of architectural knowledge. However, SO's wealth of knowledge is predominantly represented as unstructured text, which poses challenges for searching and reusing architecture-specific knowledge (e.g., architectural solutions). The abstract nature of architectural concepts exacerbates this issue, making keyword-based search methods insufficient for effectively locating relevant solutions. Recently, semantic-based approaches leveraging deep learning and advanced information retrieval techniques have demonstrated promising results in searching and reusing software artifacts, such as code. These approaches could serve as a foundation for developing innovative tools that better support OSS developers in searching and reusing architectural solutions from Q\&A sites like SO. Tool designers could also develop custom tools or browser extensions tailored for OSS development. These tools could leverage Natural Language Processing (NLP) techniques to align developers' design concerns with relevant architectural solutions, enabling more refined and targeted searches on SO. Additionally, the tools can prioritize solutions based on critical factors such as scalability, maintainability, or security, effectively bridging the gap between SO's vast architectural knowledge and the specific needs of OSS developers. By facilitating more effective search and reuse of architectural solutions, such tools would empower developers to address complex architectural issues more efficiently, ultimately advancing OSS development practices.

\textbf{Architectural tools for adapting and integrating architectural solutions from SO}. The results from RQ3 highlight that OSS developers employ various ways to adapt architectural solutions in their projects. Among these, \textit{Adapting the architectural solutions to fit in the project context} emerges as one of the most frequently used ways in OSS development. This reinforces the conclusion of Bedjeti \textit{et al}. \cite{bedjeti2017modeling} that design contexts are essential for guiding the architecture design of a system. OSS developers must tailor architectural solutions from SO to fit the specific needs of their project contexts. These insights highlight that identifying architectural solutions is merely the starting point of their effective utilization. Once relevant solutions are retrieved from SO, significant effort and domain knowledge are required to adapt and refine these solutions for effective integration into the desired design contexts. Moreover, the results of RQ5 shed light on the challenges developers face when using these solutions. Notable challenges include \textit{significant time required to adapt architectural solutions from SO to address specific design concerns in OSS development} and \textit{architectural solutions from SO lacking the appropriate design context for the given OSS project} (see Table \ref{ChallengesFacedWhenUtilizingArchitecturalSOlutions}). These challenges are often exacerbated by the absence of dedicated tools and approaches to efficiently adapt and integrate architectural solutions into target projects. Such findings emphasize the potential for developing tools that automate or streamline the solution adaptation process. Tool designers could leverage the results of this study (specifically from RQ4 and RQ5) to provide architectural tools that assist developers in adapting and integrating architectural solutions from SO to their project-specific contexts. To maximize practical applicability, these tools could be designed as plugins for Integrated Development Environments (IDEs), such as PyCharm, Eclipse, and NetBeans. By integrating seamlessly into existing OSS development workflows, these tools could help bridge the gap between general architectural solutions and project-specific needs.
\color{black}

\section{Threats to Validity}\label{ThreatsValidity}
\textcolor{black}{This section discusses the threats to the validity of our study and the measures adopted to mitigate them.}

\subsection{Construct Validity}
\textcolor{black}{\textbf{Limited scope of query terms}. In this study, we used the search term “architect*” to identify commits and issues discussing architectural design (see Section \ref{CollectingDataforMiningStudy}). While this approach focused on capturing the most relevant content about architecture, it may have excluded discussions that used alternative terms, such as “design”. To address this, we conducted a pilot study showing that OSS developers often use “design” in commit messages or issue comments to refer to programming contexts unrelated to architectural design. For example, the term “design” was used in a Uint32Array design\footnote{\url{https://github.com/saghul/txiki.js/issues/369}} context, which was implementation-focused rather than architectural. To further mitigate this threat, we considered broadening the query to include “design”, but this significantly increased the number of irrelevant results. A pilot query incorporating “design” retrieved 20,000 results (2,000 commits and 18,000 issues). We then calculated a representative sample size of these 2K commits and 18K issues. With a 95\% confidence level and 3\% margin of error, the representative sample size calculated is 697 commits and 1,008 issues. Afterwards, we randomly selected 1,705 (697 commits and 1008 issues) from the 20,000 commits and issues and manually checked them for calculating how many commits and issues we might have missed due to limiting the search to the “architect*” terms during the search of commits and issues. Specifically, the first author labelled the 1,705 commits and issues to determine which of the issues and commits discussed about architectural design and \textcolor{black}{reused} architectural solutions from SO to solve such problems. The second author checked and validated the labeling results. The disagreements were resolved in a meeting to improve the reliability of the labeling results. Based on our manual labeling, we found that out of the 1,705 commits and issues, only 32 were relevant (i.e., the true positives). Of these, 14.3\% (i.e., 4 out of 32) did not contain “architect*” terms, while 85.7\% (i.e., 28 out of 32) contained “architect*” terms. Therefore, we admit that we might have missed a certain number of relevant commits and issues (i.e., 14.3\%) that do not contain “architect*” terms. This limitation impacts the completeness of our findings, as some architectural discussions may have been overlooked. Moreover, we relied on manual searches rather than automated approaches, such as Machine Learning (ML), which poses potential threats, including incomplete data retrieval, human error, and biases in identifying relevant discussions. While the use of ML to automate data retrieval and filtering is indeed a valuable approach, we opted for a manual method to ensure accuracy and thoroughness in data collection and filtering. This allowed us to closely examine the data and ensure its relevance to the study's objectives. However, we admit that there might still be some relevant discussions that were inadvertently overlooked due to the limitations of manual searches, such as subtle or indirect references to architectural concepts. To address these limitations, future work will focus on improving query strategies through NLP techniques, ML classifiers, and synonym expansion to capture discussions using alternative terms while maintaining precision. Cross-referencing multiple data sources, such as SO tags or GitHub issue labels, could enhance recall and reduce irrelevant results. Furthermore, incorporating automated approaches will improve the efficiency and effectiveness of the data collection method, enabling a more comprehensive exploration of architectural discussions in OSS development and SO posts.} 

\textcolor{black}{\textbf{Limitations of using mostly semi-closed questions}. Another potential threat to the construct validity of this study is the predominance of semi-closed questions (e.g., SQ7, SQ8) in the survey, rather than open-ended questions. This design choice may limit the depth and richness of the responses collected from participants. However, as noted by Reja \textit{et al}. \cite{reja2003open}, open-ended questions have several disadvantages compared to closed-ended or semi-closed questions. Specifically, they can increase the time required to complete the survey, potentially leading to lower participation rates. Additionally, they may lead to vague, incomplete, or skipped responses. To mitigate these issues, we primarily utilized semi-closed questions, which provide predefined multiple-choice options alongside an “Other” field. This design allowed respondents to supply alternative answers when the provided options did not fully apply. By incorporating an “Other” field, we aimed to balance the need for structured responses with the flexibility to capture diverse perspectives, while also minimizing participant fatigue and promoting higher response quality.} 

\textcolor{black}{\textbf{Issues in tailoring the survey}. As discussed in Section \ref{CollectingSurveyData}, we presented survey participants with a randomly selected set of candidate answers categorized based on findings from the mining study. While this approach helped focus responses, it also had limitations, as not all possible categories were included, potentially narrowing the range of participant insights. We intentionally limited the categories presented for two primary reasons: (1) to uncover new insights, such as previously unidentified categories of architectural problems or challenges, and (2) to prevent participants from being overly influenced by the mining study results. To enhance clarity and relevance, personalized emails with tailored versions of the questionnaire were sent to developers, linking the architectural solutions they had used on SO to corresponding commits or issues (see Section \ref{CollectingSurveyData}). This customization was guided by two considerations: minimizing misunderstandings and adhering to best practices from prior studies~\cite{martinez2021did, li2020qualitative, li2021understanding}. Despite these efforts, potential biases were introduced. For example, providing candidate answers based on the mining study and targeting developers tied to specific mined data points may have overlooked other decisions or broader experiences (e.g., \textcolor{black}{reused} architectural solution) that participants might have wanted to share. To address this, we incorporated an “Other” field in the semi-closed multiple-choice questions, allowing participants to offer additional responses or insights beyond the predefined options. This measure aimed to reduce constraints and biases, providing participants with flexibility to share their experiences comprehensively
}

\textcolor{black}{\textbf{Limited cross-validation of survey responses to answer the RQs}. While the first three research questions (RQ1, RQ2, and RQ3) are based on data from both GitHub and survey responses, RQ4 and RQ5 rely exclusively on the survey data (SQ9 and SQ11, respectively). This reliance on a single data source for RQ4 and RQ5 may be considered a limitation in terms of data variety and robustness, potentially introducing validity threat due to the lack of cross-validation. To address this concern, we conducted a pilot study, which revealed that OSS developers rarely mentioned characteristics of architectural solutions (relevant to RQ4) or challenges faced (relevant to RQ5) in commit messages or issue comments. In the formal study, we revisited these sources for any data pertinent to RQ4 and RQ5 but did not find additional relevant information, confirming the survey as the primary data source. While this single-source approach may still be seen as a limitation, we believe the steps taken, particularly the pilot study and revisiting GitHub data, have partially mitigated this validity threat.}

\textcolor{black}{\textbf{Lack of direct validation between the two research methods}. As clarified in Section \ref{Methodology} and Section \ref{CollectingSurveyData}, we conducted an online survey with practitioners contributing to OSS systems, aiming to complement, not validate, the results of the mining study by gathering additional insights. Since the survey was not designed to confirm or refute the findings of the mining study, the lack of direct validation between the two methods could be seen as a threat. This raises the possibility that the results from one method may not align with the other, potentially impacting the study's complementary nature. Without direct validation, uncertainties may arise regarding the accuracy or completeness of the findings. However, we designed the survey to capture a wide range of perspectives on the mining study's results. By doing so, we ensured that the survey added new dimensions to the findings, helping to deepen our understanding of the utilization of architectural solutions in OSS development and the challenges faced by developers.}

\subsection{External Validity}
\textcolor{black}{\textbf{Exclusive focus on GitHub and SO as data sources, \textcolor{black}{and limited participant diversity}}. Our study examined commits and issue reports from GitHub and architectural solutions discussed on SO. While GitHub and SO are prominent repositories of software development knowledge and have been widely used in prior software engineering research (e.g., \cite{chen2024empirical}), this exclusive focus may limit the generalizability of our findings to other developer communities. These platforms may not fully represent the diversity of activities related to the use of architectural solutions in software development. To address this limitation, we recommend that future research can expand the scope by employing alternative methods, such as interviews or case studies, and incorporating additional platforms as data sources. Examples include SourceForge, GitLab, and Reddit, which might reveal different insights about how architectural solutions from Q\&A sites are adopted in various contexts. Moreover, our survey targeted developers who are active OSS contributors on GitHub and who utilize architectural solutions from SO. Consequently, our findings may not generalize to other developer populations, such as those working on closed-source projects or in different ecosystems. Future studies should extend both the mining study and survey, offering a broader understanding of architectural solution usage across diverse development contexts and regions.}

\subsection{Reliability} 
\textcolor{black}{\textbf{The issues surrounding the manual data filtering}. One threat to reliability in this study arises from the manual data filtering process, where we used a limited sample size to define the inclusion/exclusion criteria. Specifically, during the initial pilot stage of data filtering (see Section \ref{CollectingDataforMiningStudy}), we used a sample of 10 issues and 10 commits to establish our inclusion/exclusion criteria (see Table \ref{InclusionExclusionCriteria}). This limited sample size may pose a reliability risk, potentially introducing bias and affecting the broader generalizability of data filtering results. To mitigate this, the first author followed an iterative process to refine our criteria throughout the former data filtering process, revisiting and adjusting the inclusion/exclusion criteria as the number of commits and issues increased. When commits or issues were unclear, and the first author encountered ambiguity during filtering, meetings with the other two authors were held to discuss and resolve any discrepancies. This iterative and collaborative approach helped to enhance the consistency and reliability of our filtering criteria and results.} 

\textcolor{black}{\textbf{The issues surrounding the manual data analysis}. Another notable threat to reliability relates to the manual analysis of the selected commits and issues. Specifically, we piloted the process using a small sample of 10 issues and 10 commits. Manual analysis is inherently susceptible to personal bias due to variations in interpretation or oversight. Additionally, the limited sample size raises concerns about reliability, potentially impacting the broader generalizability of our findings. To address this, we implemented a rigorous protocol to standardize the manual analysis process. This included predefined qualitative analysis guidelines, cross-validation among multiple researchers, and the application of two widely used techniques in software engineering: open coding and constant comparison~\cite{seaman1999qualitative}, as well as descriptive statistics~\cite{easterbrook2008selecting}, as detailed in Section \ref{Methodology} and Section \ref{CollectingDataforMiningStudy}. Pilot data coding was conducted before the formal data analysis to further mitigate these issues. During the pilot phase, the first author selected a random sample of 10 commits and 10 issues and encoded the extracted data in alignment with the objectives of each RQ (see Table \ref{RelationshipDataAnaysisMethodandRQandSQ} and Section \ref{Introduction}). This pilot phase involved an iterative process where the first author continuously refined the codes, concepts, and categories by revisiting the extracted data. To address any ambiguities, the first author held several in-person meetings with the second author. The pilot data coding results, including concepts and categories, were reviewed and validated by the second author. Disagreements were resolved collaboratively using the negotiated agreement approach~\cite{campbell2013coding}, thereby enhancing the reliability of the pilot data analysis. While the pilot data analysis involved a small dataset, the subsequent formal data analysis followed the same procedures in the pilot data analysis to enhance the reliability of the data analysis results and ensure robust findings. With these measures, the threats associated with the manual data analysis have been partially mitigated.}

\section{Conclusions}\label{Conclusions}
While prior studies have extensively explored the use of source code from Q\&A sites in OSS development, there has been little focus on how developers utilize architectural solutions, a type of high-level artifacts, from these platforms. This study addresses this gap by providing a comprehensive overview of how architectural solutions from SO are adopted and \textcolor{black}{reused} in OSS development. Specifically, we investigated the architectural problems addressed by these solutions, the characteristics of these solutions and their descriptions, the ways the solutions are \textcolor{black}{reused}, and the challenges practitioners face when \textcolor{black}{reusing} the solutions.

Our findings reveal that practitioners use architectural solutions from SO to address a broad range of architectural problems (15 categories) in OSS development, leveraging architectural solutions across seven identified categories. Practitioners adopt five methods when \textcolor{black}{reusing} these solutions, with \textit{converting ideas in the architectural solutions into code} and \textit{adapting the architectural solutions to fit in the project context (design context)} being the most prevalent. Notably, practitioners prioritize architectural solutions that include code samples and design context descriptions. However, they face significant challenges, with the most frequently reported being the \textit{considerable time required to adapt architectural solutions from SO to address OSS design concerns}.

This study offers implications for researchers and practitioners in the software architecture community. Dedicated tools and approaches, such as IDE plugins, can be developed to address these challenges and streamline the adaptation of architectural solutions from SO. These tools should account for the adaptation time and ensure alignment with the design contexts of OSS projects. \textcolor{black}{Additionally, while this study focused on reusing architectural solutions from SO in GitHub, future research could expand this scope and investigate additional data sources and techniques for improving architectural knowledge reuse. For example, tool designers could explore the development of hybrid search engines that integrate architectural solutions from Q\&A fora with scientific literature and formal standards. Such systems could enhance architectural knowledge reuse by offering developers access to both community-driven insights and research-backed solutions.} Moreover, our findings provide actionable insights for Q\&A site owners and users. For instance, SO could enhance its support for architectural solutions by introducing mechanisms such as voting on quality aspects (e.g., performance and security) of solutions or enabling tagging of solutions (rather than tagging questions alone), making it easier for practitioners to discover, evaluate, and utilize relevant solutions. \textcolor{black}{Furthermore, while the number of votes and the presence of an accepted answer indicator are often seen as important cues that may influence OSS developers’ selection and reuse of solutions on SO, these factors did not emerge as key findings in our study. Future work could empirically examine the extent to which such community signals, like vote counts, affect developers’ decisions to reuse architectural solutions. Gaining a deeper understanding of these cues could support the development of recommendation systems that more effectively align with developers' decision-making behaviors in practice.}

\section*{Acknowledgments}
This work has been partially sponsored by the NSFC under Grant No. 62172311 and the Major Science and Technology Project of Hubei Province, China under Grant No. 2024BAA008. The numerical calculations in this paper have been done on the supercomputing system in the Supercomputing Center of Wuhan University. The authors would also like to thank all the participants in the practitioner survey.

\bibliographystyle{ieeetr}
\bibliography{ref}

\end{sloppypar}
\end{document}